\documentclass[aps,prl,reprint,showpacs,superscriptaddress,longbibliography]{revtex4-2}

\usepackage[dvipdfmx]{graphicx}
\usepackage{bm,ulem}
\usepackage{hyperref}
\usepackage{IEEEtrantools}
\usepackage{pgffor}
\usepackage{pdfpages}
\usepackage{diagbox}
\usepackage{lineno}
\usepackage{booktabs}

\usepackage{mathrsfs}
\usepackage{natbib}
 \usepackage{amsfonts}
\usepackage{physics}
\usepackage{float}
\usepackage{siunitx}
\usepackage{xcolor}
\usepackage{comment}
\hypersetup{
setpagesize=false,
colorlinks=true,
linkcolor=blue,
citecolor=red,
}

\makeatletter
\AtBeginDocument{\let\LS@rot\@undefined}
\makeatother

\def\supplementfilename{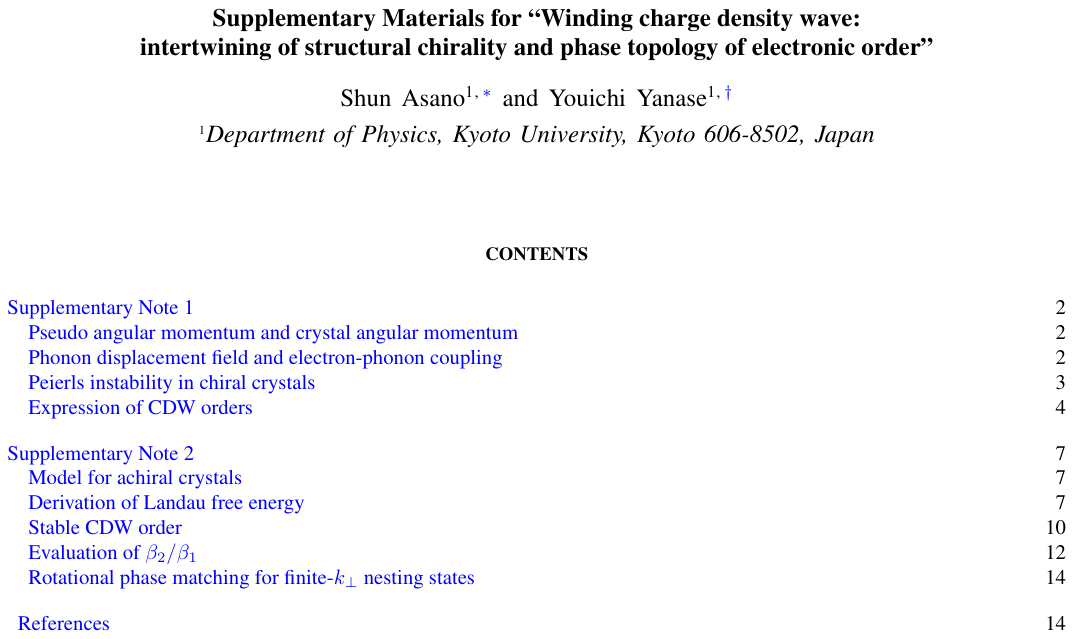}

\pdfximage{\supplementfilename}
\def\numbersupplementpages{\the\pdflastximagepages}

\newif\ifarXiv
\arXivtrue

\restylefloat{table}

\begin{document}

\title{Winding charge density wave: \\intertwining of structural chirality and phase topology of electronic order}

\author{Shun Asano}
\email{asano.shun.57x@st.kyoto-u.ac.jp}
\affiliation{Department of Physics, Kyoto University, Kyoto 606-8502, Japan}

\author{Youichi Yanase}
\email{yanase@scphys.kyoto-u.ac.jp}
\affiliation{Department of Physics, Kyoto University, Kyoto 606-8502, Japan}

\begin{abstract}
 We propose a class of chiral charge density waves (CDWs), dubbed winding CDWs, that exhibit macroscopic chirality despite a single ordering wavevector. In screw-symmetric chiral crystals, chiral phonons drive a Peierls instability that selects a definite crystal angular momentum channel, thereby endowing the CDW with an integer azimuthal phase winding dictated by the selection rule governing electron–phonon coupling. We further extend this framework to achiral crystals with discrete rotational symmetry and demonstrate that spontaneous symmetry breaking stabilizes a winding CDW with either handedness, realizing an achiral-to-chiral phase transition. Our results reveal a fundamental link between the geometry of chiral structures and the phase topology of electronic orders.
\end{abstract}

\maketitle

\textit{Introduction}.---Non-centrosymmetric materials, lacking a center of inversion, have long fascinated researchers across various disciplines~\cite{ErnstBauer2012,Sungkit2014annurevphys,Burkov2018annurevphys,Tokura2021ChemRev}.
Among them, chiral structures, which cannot be superimposed onto their mirror images, have been a particularly prominent class owing to their unique properties~\cite{Bousquet2025JPhys,Barron2004,ZhuHanyu2024NatMater,Ni2021SmatMat,Bloom2024ChemRev,Yan2024AnnualRevMatRes}. Chiral crystals can host microscopic excitations that themselves possess chiral character. An example is the chiral phonon, a lattice vibration with circular atomic motion that carries angular momentum~\cite{ChenHao20192DMater,Wang2024NanoLett,Juraschek2025NatPhys}. Recent studies have further revealed that structural chirality can shape the topological properties of quasiparticles, establishing chiral crystals as a platform where crystalline geometry and band topology are intertwined~\cite{Zhong-Ke2024AdvFunctMater,Tzhang2025arxiv,Chang2018NatMater,Krieger2024NatComm,Zhang2023NanoLett,LiuYizhou2021NatMater}.

Chirality, however, is not confined to static crystal structures. Electronic, magnetic, and structural orders can acquire handedness through external fields or spontaneous symmetry breaking, even in structurally achiral crystals.~\cite{Romao2024ACSnano,Zheng2025Science,Fava2025PRL,Xu2020Nature,Hayashida2021JACS,Ishitobi2026PRL}. A representative class is the charge density wave (CDW), a collective electronic order commonly driven by the Peierls mechanism, namely Fermi-surface nesting and electron–phonon coupling (EPC)~\cite{Gruner1988RevModPhys,Monceau2012AdvPhys,Xuetao2015PNAS}. Known chiral CDWs typically rely on multiple-$Q$ ordering patterns with mutually rotated wavevectors~\cite{Ishioka2010PRL,Yang2022PRL,Kim2024NatPhys,Zhao2023NatComm,Zhang2024npj}, while transverse Peierls transitions (PT) have recently been proposed as another route to chiral CDWs through transverse lattice distortions~\cite{Luo2023PRX,Yang2025NatComm}.

Although these studies demonstrate that chirality can emerge as a collective ordered state, they leave unresolved whether and how the  topology of such an order is related to static crystal structures. This raises the fundamental question of whether structural chirality, beyond the mere absence of improper spatial symmetries, provides a geometric framework that governs the winding of electronic order, and whether an analogous rotational-symmetry framework can generate chirality spontaneously in achiral crystals.

\begin{figure}
    \centering
    \includegraphics[width=0.9\linewidth]{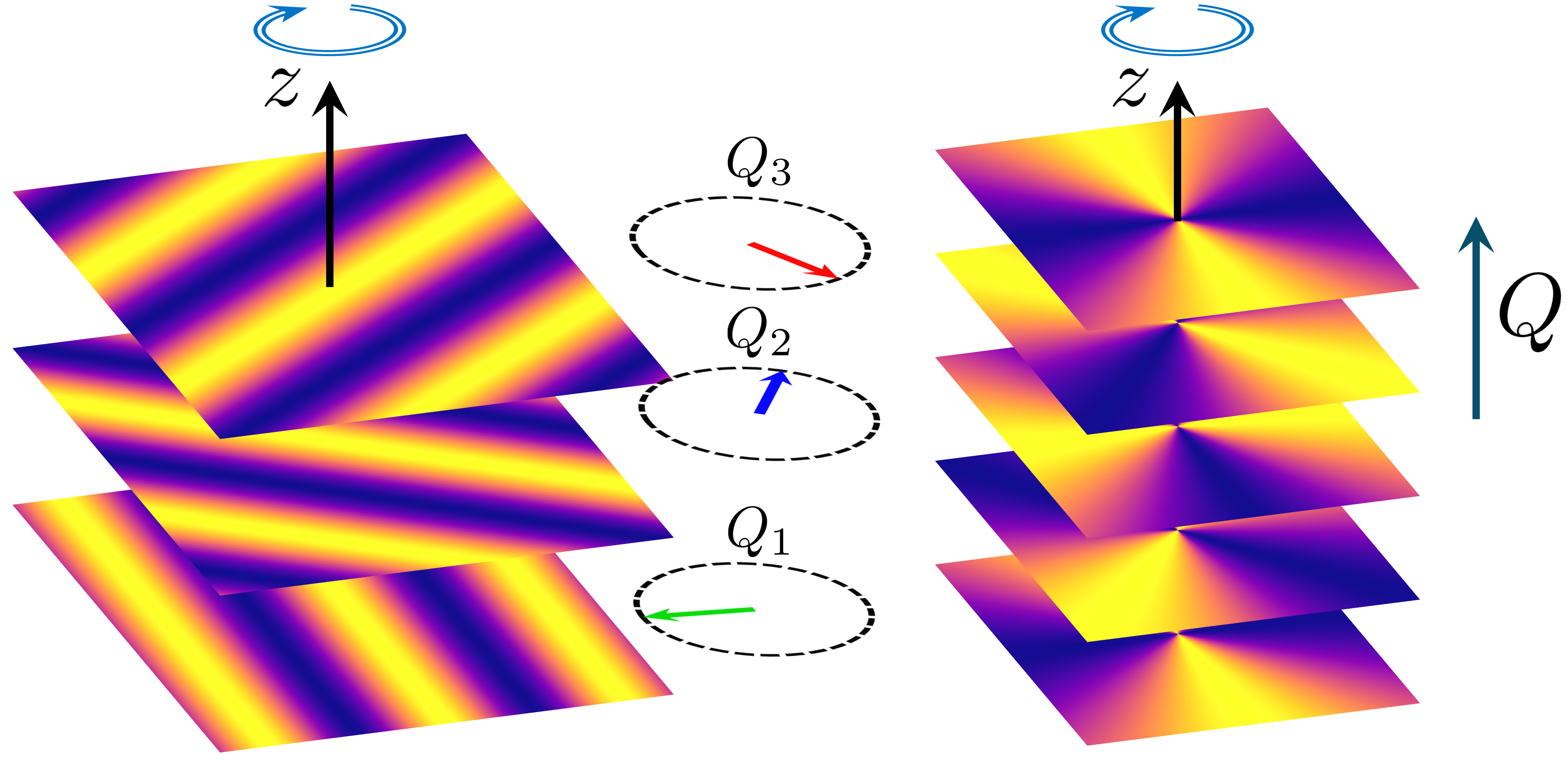}
    \caption{Schematic illustrations of chiral CDW orders. (Left) Chiral CDW characterized by triple-$Q$ nesting vectors. (Right) Winding CDW with a single-$Q$ wavevector, yet exhibiting chirality. Shading denotes charge density variations.}
    \label{fig1}
\end{figure}

Here we demonstrate an unrecognized class of chiral electronic order in both chiral and achiral crystals. Unlike chiral CDWs discussed so far which require multiple $Q$ related by mutual rotations~\cite{Ishioka2010PRL,Yang2022PRL,Kim2024NatPhys,Zhao2023NatComm,Zhang2024npj,ShaoSen2026PRL}, this order develops a macroscopic chiral structure even though it is uniaxial, i.e., characterized by a single $Q$ [Fig.~\ref{fig1}]. The key ingredient is crystal angular momentum (CAM) conservation in electron–phonon coupling: the CAM difference of nested electronic states is transferred to the azimuthal winding of the CDW. Consequently, this CDW order is characterized by an integer winding number, analogous to the topological charge of vortex waves carrying orbital angular momentum~\cite{Lloyd2017RevModPhys,Bliokh2017PhysRep}.
We therefore refer to it as a winding CDW to distinguish it from conventional chiral CDWs. In the first part of this Letter, we show that a transverse PT in screw-symmetric chiral crystals naturally leads to the formation of winding CDWs. In the second part, we extend our formulation to achiral crystals with discrete rotational symmetry and demonstrate that spontaneous symmetry breaking can stabilize winding CDWs, thereby realizing an achiral-to-chiral phase transition.

\textit{Chiral phonons and axial phonons}.---Following Ref.~\cite{Juraschek2025NatPhys}, we distinguish chiral phonons, which break improper rotational symmetries~\cite{Ueda2023Nature,Chen2022NanoLett,Ishito2023NatPhys,Chen2021NanoLett,Zhang2025NatPhys}, from axial phonons, which carry nonzero real and/or pseudo angular momentum without necessarily being chiral~\cite{Hanyu2018Science,Zhang2015PRL,Zhang2014PRL}. This distinction is useful here, because the present framework applies to both chiral and achiral crystals. Furthermore, although not mentioned in Ref.~\cite{Juraschek2025NatPhys}, even for degenerate phonon bands, the linearly polarized modes can be recast, by choosing an appropriate basis, as a pair of circularly polarized axial phonons carrying opposite angular momenta~\cite{Hernandez2023SciAdv,Juraschek2017PRR,WuFangling2023NatPhys,David2024PNAS}.

While circular motion generates mechanical (i.e., real) angular momentum, pseudo angular momentum (PAM) is a physical quantity defined by the eigenvalue of an $n$-fold rotation operator $\hat{C}_n$ or screw operator $\hat{S}_n$, involved in selection rules 
governing chiral (or axial) phonon absorption/emission~\cite{Zhang2024PRR,Bozovic1984PRB,Streib2021PRB,wang2024arxiv,Tateishi2025JPSJ}. Let us consider $\hat{C}_n$-symmetric crystals and phonon modes along the corresponding high-symmetry line in reciprocal space, such as the $\Gamma$–A line considered in this work. Because the wave vector $\bm{q}$ satisfies $\hat{C}_n\bm{q}=\bm{q}+\bm{G}$ with a reciprocal lattice vector $\bm{G}$, the phonon Bloch function $\bm{u}_{\bm{q}}$ becomes an eigenvector,
\begin{equation}
\hat{C}_n \bm{u}_{\bm{q}} = e^{-2\pi i \,l_{\mathrm{ph}}/n} \, \bm{u}_{\bm{q}} ,
\end{equation}
where $l_{\mathrm{ph}}$ is the phonon PAM and takes integer values modulo $n$.
Similarly, if the crystal possesses screw symmetry $\hat{S}_n=\hat{C}_n \hat{T}_{c/n}$, where $\hat{T}_{c/n}$ is the translation operation along the rotational axis (here taken as the $z$-axis with a lattice constant $c$), the phonon mode satisfies
\begin{equation}
\hat{S}_n \bm{u}_{\bm{q}}
=e^{-2\pi i \,l_{\mathrm{PAM}}/n} \, \bm{u}_{\bm{q}}
\equiv
e^{-(2\pi i \,l_{\mathrm{ph}}/n+ i q_zc/n)} \, \bm{u}_{\bm{q}} .
\end{equation}
In the case of screw symmetry, an additional phase $q_zc/n$ arises due to the translational operator contained in $\hat{S}_n$, so that the phonon PAM $l_{\mathrm{PAM}}=l_{\mathrm{ph}}+q_zc/2\pi$ becomes fractional values dependent on $\bm{q}$. However, $l_{\mathrm{ph}}$ is a purely rotational part of the phonon PAM and represents a contribution in common with the $\hat{C}_n$-symmetric case~\cite{Tzhang2025arxiv}. Therefore, the quantized integer $l_{\mathrm{ph}}$ is hereafter called the CAM for both $C_n$ and $S_n$-symmetric crystals~\cite{Tateishi2025JPSJ}.

Similarly, the electronic PAM can also be defined for electronic Bloch states at high symmetry lines:
\begin{equation}
\hat{C}_n \psi_{\bm{k}} = e^{-2\pi i \,m_{\mathrm{PAM}}/n} \psi_{\bm{k}}. 
\end{equation} 
In general, when the electron spin is taken into account, the spin space is simultaneously rotated by the rotational operator. As a result, the electronic PAM includes a contribution from the electron spin and should be treated within the double-group representation~\cite{Bradley2009,asano2026arxiv}. In this Letter, we focus on spinless electronic states for brevity, or on situations where the spin degree of freedom is not involved in the relevant interactions. Therefore, we exclude the electron spin from the electronic PAM. In this case, as for phonons, the eigenvalues of the $\hat{C}_n$ operator $m_{\mathrm{el}}$ take integers, while for the $\hat{S}_n$ operator, an additional phase $k_zc/2\pi$ appears.

The ordered states considered below are dominated by the scattering process parallel to the rotational or screw axis. Moreover, in the electron–phonon scattering process, the translational contributions $q_zc/2\pi,  \, k_zc/2\pi$ that appear in phononic and electronic PAM cancel in the conservation law. This is because the conservation of crystal momentum holds. As a result, the conservation law for the EPC can be reduced to that of total CAM~\cite{Zhang2024PRR,Bozovic1984PRB,Streib2021PRB,wang2024arxiv,Tateishi2025JPSJ}:
\begin{equation}
m_{\mathrm{el}} + l_{\mathrm{ph}}=m'_{\mathrm{el}}
\quad \mathrm{mod}\; n,  
\label{CAM_selection}
\end{equation}
which enables us to classify unconventional CDWs.

\begin{figure*}[t]
    \centering
    \includegraphics[width=\linewidth]{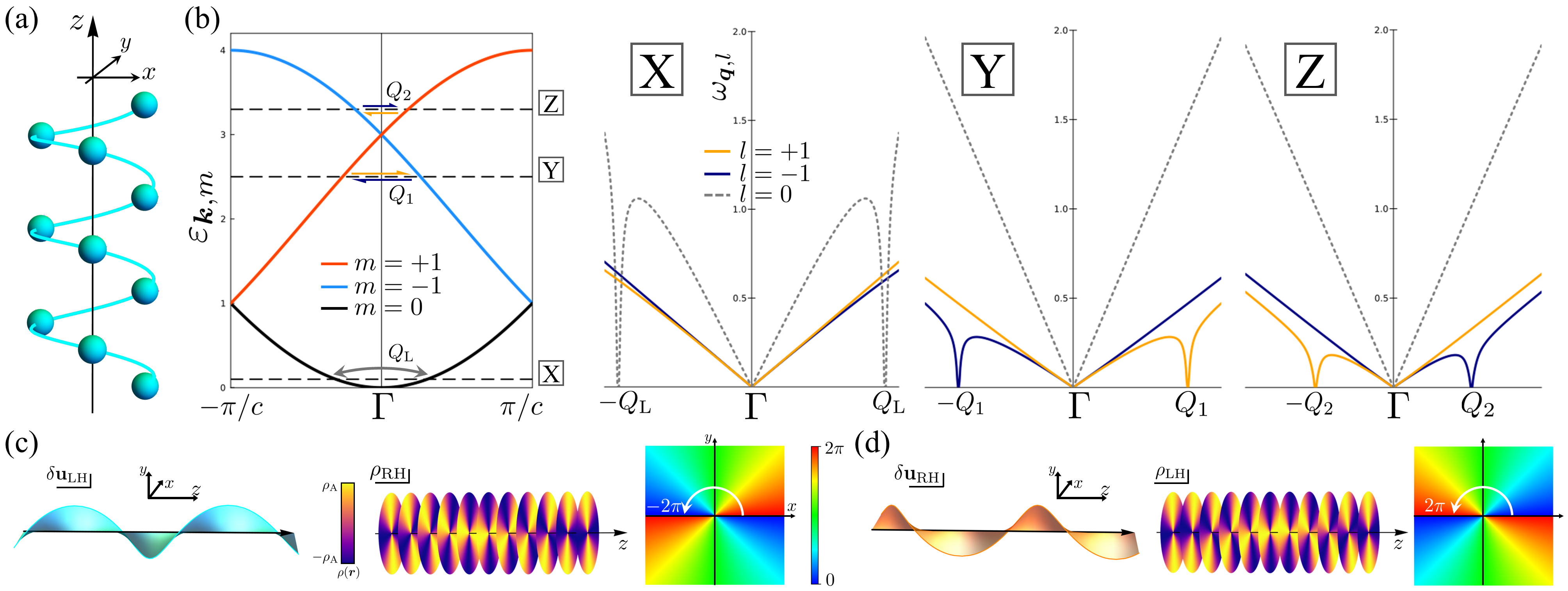}
    \caption{Winding CDW in the chiral crystal. (a) An illustration of $\hat{S}_3$-symmetric crystal. (b) (Left) spinless electronic bands measured from band bottom with CAM $m=0, \pm1$. The dashed lines denote chemical potentials for different cases of Peierls instability. (Right) The phonon branches at the transition temperature for each instability. Case X corresponds to longitudinal PT, while Y and Z correspond to right-handed (RH) and LH winding CDW, respectively. (c) The RH winding CDW and accompanying LH lattice displacement arising in case Y. The inset shows the in-plane CDW phase distribution, $\arg \rho(\mathrm{r})$. (d) The LH winding CDW and accompanying RH lattice displacement arising in case Z.}
    \label{fig2}
\end{figure*}

\textit{Formulation in chiral crystals}.---As a typical example, we consider $\hat{S}_3$-symmetric chiral crystals and mainly focus on left-handed (LH) crystals as shown in Fig.~\ref{fig2}(a)~\cite{SM1}. Therefore, in the following, CAM is defined as $m=0,\pm1$ for electrons and $l=0,\pm1$ for phonons. We study conventional $2k_{\rm F}$-instability as a representative and well-established setting and assume that the electronic bands possess the well-nested property with $\bm{Q}=(0,0,Q)$ along the rotational $z$ axis (i.e., along the $\Gamma$–A line)~\cite{Gruner1988RevModPhys,Monceau2012AdvPhys}.

The minimal Hamiltonian satisfying the conservation law, Eq.~\eqref{CAM_selection}, becomes
\begin{align}
\mathcal{H}=\mathcal{H}_{\mathrm{el}}+\mathcal{H}_{\mathrm{ph}}+\mathcal{H}_{\mathrm{ep}}^{(\mathrm{L})}+\mathcal{H}_{\mathrm{ep}}^{(\mathrm{T})}.
\label{Hamiltonian}
\end{align}
Here, $\mathcal{H}_{\mathrm{el}}=\sum_{\bm{k},m}\xi_{\bm{k},m}\hat{c}_{\bm{k},m}^{\dagger}\hat{c}_{\bm{k},m}$ and $\mathcal{H}_{\mathrm{ph}}=\sum_{\bm{q},l}\hbar\omega_{\bm{q},l}\hat{b}_{\bm{q},l}^{\dagger}\hat{b}_{\bm{q},l}$ are the electron and phonon parts, respectively, with the electron creation (annihilation) operator, omitting the spin index, $\hat{c}_{\bm{k},m}^{\dagger}$ ($\hat{c}_{\bm{k},m}$), those of the phonon $\hat{b}_{\bm{q},l}^{\dagger}$ ($\hat{b}_{\bm{q},l}$), the electron dispersion $\xi_{\bm{k},m}$, and the phonon dispersion $\omega_{\bm{q},l}$. The EPC Hamiltonian is written as follows:
 \begin{align}
    \mathcal{H}_{\mathrm{ep}}^{(\mathrm{L})}&=\dfrac{1}{\sqrt{N_{\mathrm{i}}}}\sum_{\bm{k},\bm{q},m}iqg_{\mathrm{L}}\zeta_{\bm{q},0}\,\hat{c}_{\bm{k}+\bm{q},m}^{\dagger}\hat{c}_{\bm{k},m}(\hat{b}_{\bm{q},0}+\hat{b}_{-\bm{q},0}^{\dagger}) ,
 \\
    \mathcal{H}_{\mathrm{ep}}^{(\mathrm{T})}&=\dfrac{1}{\sqrt{N_{\mathrm{i}}}}\sum_{\bm{k},\bm{q},m}\sum_{l=\pm1}iqg_{\mathrm{T}}\zeta_{\bm{q},l}\,\hat{c}_{\bm{k}+\bm{q},m+l}^{\dagger}\hat{c}_{\bm{k},m}(\hat{b}_{\bm{q},l}+\hat{b}_{-\bm{q},-l}^{\dagger}),
\end{align}
where $\zeta_{\bm{q},l}=\sqrt{\hbar/({2M_{\mathrm{i}}\omega_{\bm{q},l}})}$, $N_{\mathrm{i}}$ and $M_{\mathrm{i}}$ are the number and mass of ions, respectively, and $g_{\mathrm{L}}$ ($g_{\mathrm{T}}$) is the coupling constant for the longitudinal (transverse) mode. Since we focus on the $\Gamma$–A line, we can reduce this model to states in the vicinity of $\bm{k}=(0,0,k_z)$ and $\bm{q}=(0,0,q_z)$. In chiral crystals, the absence of inversion ($\mathcal{P}$) and mirror symmetries lifts the degeneracy between opposite angular momenta. While time-reversal ($\mathcal{T}$) symmetry enforces $\omega_{-q_z,-l}=\omega_{q_z,l}$, the dispersion of chiral phonons satisfies $\omega_{q_z,+1} \neq \omega_{q_z,-1}$. Similarly, the electronic bands satisfy $\xi_{k_z,+1} \neq \xi_{k_z,-1}$. As an example, Fig.~\ref{fig2}(b) shows the electron bands $\xi_{k_z,m}=-t \cos (k_z c/3 + 2m\pi/3)-\mu$ in the LH crystal.

The EPC Hamiltonian is decomposed into the $l=0$ and $l=\pm1$ sectors. The former, $\mathcal{H}_{\mathrm{ep}}^{(\mathrm{L})}$, describes scattering processes that conserve the electronic CAM, whereas the latter, $\mathcal{H}_{\mathrm{ep}}^{(\mathrm{T})}$, describes processes in which the electronic CAM changes by $\pm1$. The $l=0$ mode, although not strictly identical to the longitudinal component, corresponds to the longitudinal phonon, while the $l=\pm1$ modes correspond to the transverse phonons carrying finite angular momentum. In this sense, the two sectors correspond to the EPC for longitudinal and transverse phonon modes, respectively.

In this setting, both conventional PT driven by longitudinal phonons and transverse PT driven by transverse phonons are, in principle, possible. However, when the chemical potential intersects only the electronic bands with $m=\pm1$ [see Fig.~\ref{fig2}(b)], only phonons with $l=\pm1$ can mediate scattering processes near the nested Fermi surface due to the selection rule \eqref{CAM_selection}. Therefore, compared to the Fermi-surface instability responsible for conventional PT, that for transverse PT becomes significantly larger. Consequently, transverse PT is expected to be favored.

\begin{figure*}[t]
    \centering
    \includegraphics[width=\linewidth]{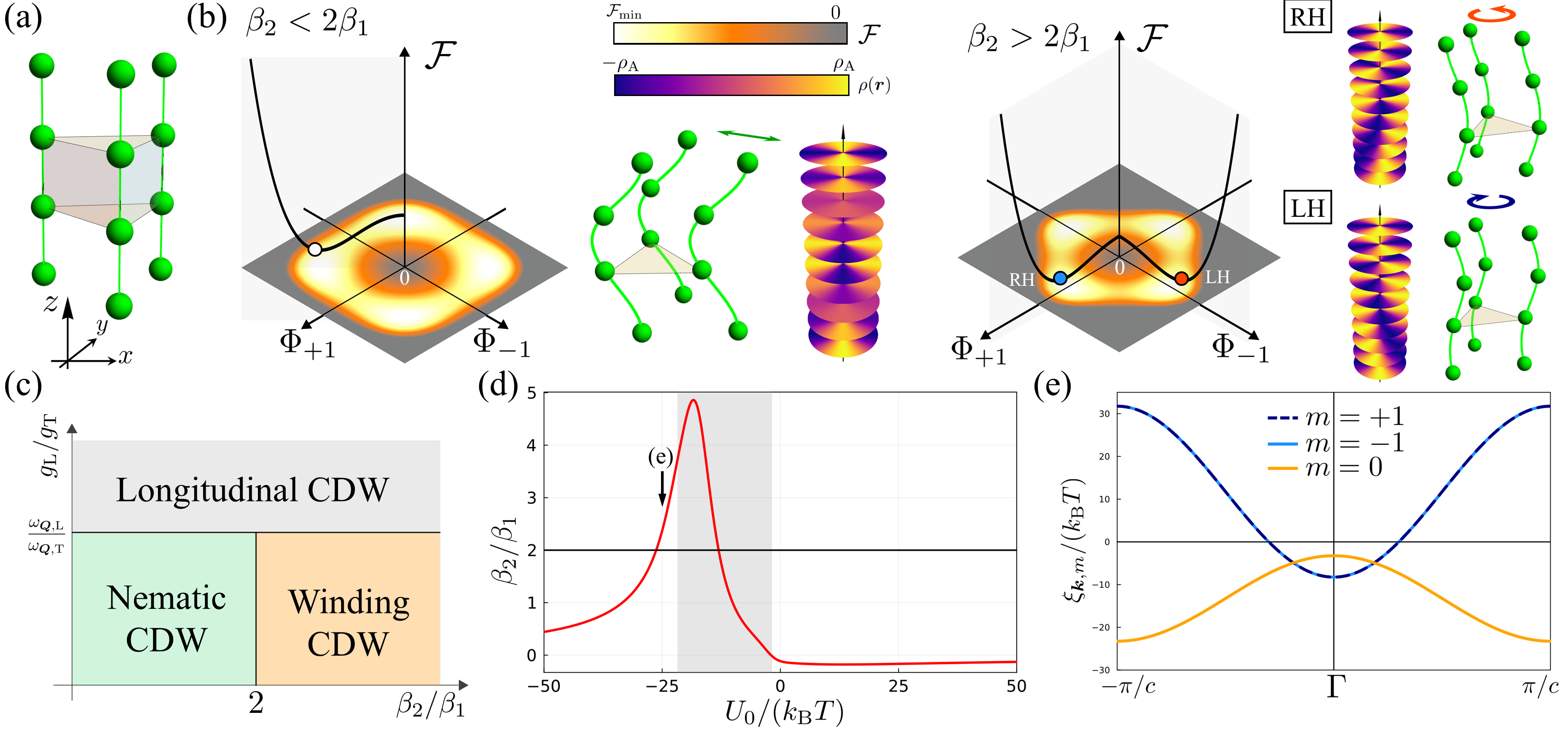}
    \caption{CDWs in achiral crystals. (a) An illustration of $\hat{C}_3$-symmetric crystals of parent phase. (b) Free energy landscape below $T_{\mathrm{C}}$ for (left) $\beta_2<2 \beta_1$ and (right) $\beta_2>2 \beta_1$, and an illustration of  (left) nematic CDW and (right) winding CDW phases at the free energy minima. (c) Phase diagram in the $g_{\mathrm{L}}/g_{\mathrm{T}} - \beta_1/\beta_2$ plane. (d) The calculated result of $\beta_2/\beta_1$ as a function of $U_0/(k_{\mathrm{B}}T)$. We set $t_0/(k_{\mathrm{B}}T)=10, t_1/(k_{\mathrm{B}}T)=-20, k_{\mathrm{F}}=0.3\pi$. In the gray region, the Fermi surface of the electron band $m=0$ exists. (e) The electron bands above $T_{\mathrm{C}}$ for $U_0/(k_{\mathrm{B}}T)=-25$ depicted in (d) by a black arrow. The electron bands $m=\pm 1$ are degenerate in the achiral parent phase.}
    \label{fig3}
\end{figure*}

\textit{Winding CDW}.---The Peierls instability occurs between $\mathcal{T}$-paired electronic states, i.e., $(-k_{\mathrm{F}},-m)$ and $(k_{\mathrm{F}},m)$ in this case. The selection rule Eq.~\eqref{CAM_selection} imposes $m=-m+l \; \mathrm{mod}\, 3$ among this instability. Therefore, the possible pairs are constrained as $(m,l)=(\pm1,\mp1)$. Note that the trivial pair $(m,l)=(0,0)$ is also allowed, it merely corresponds to the conventional PT. Accordingly, when the Fermi energy lies on the $m=\pm1$ bands, the instability is associated with the condensation of $l=\mp1$ chiral phonons. Furthermore, in chiral crystals, because the electronic bands and the chiral phonon branches are generically split, the degeneracy between the $m=\pm1$ or $l=\pm1$ sectors is lifted. As a consequence, the corresponding transition temperatures $T_{\mathrm{C},l}$ become inequivalent, $T_{\mathrm{C},+1}\neq T_{\mathrm{C},-1}$. Under this constraint, only one of the two channels undergoes condensation at the transition, resulting in the selective condensation of a phonon mode with a definite CAM $l$.

Here, we clarify the real-space structure of the electronic order below $T_{\mathrm{C}}$, i.e., CDW, by analyzing the expectation value of the density operator $\rho(\bm{r})=\langle \hat{\Psi}^\dagger(\bm{r})\hat{\Psi}(\bm{r})\rangle$, where $\hat{\Psi}(\bm{r})=\sum_{\bm{k},m}\psi_{\bm{k},m}(\bm{r})\hat{c}_{\bm{k},m}$ is the electron field operator. The electron wave function with CAM $m$ can be explicitly described as $\psi_{\bm{k},m}(\bm{r})= u_{\bm{k},m}(r_{\perp},\varphi,z)e^{ik_z z} e^{im\varphi}$ with the azimuth angle $\varphi$, the radial coordinate $r_{\perp}$, and the periodic function $u_{\bm{k},m}(r_{\perp},\varphi,z)=u_{\bm{k},m}(r_{\perp},\varphi-2\pi/3,z-c/3)$~\cite{Bozovic1984PRB}. In the ordered state, the coupling between the paired states $(k_{\mathrm{F}},m)$ and $(-k_{\mathrm{F}},-m)$ enables a macroscopically 
ordered electron density, yielding a nontrivial phase factor $e^{i(m-(-m))\varphi}=e^{i2m\varphi}$ as well as periodic modulation $e^{iQz}$. Thus, the angular dependence of the CDW is fixed by the CAM difference $\Delta m=\pm2$, and the CDW takes a nontrivial form $e^{iQz+i\Delta m \varphi}+c.c.$, as illustrated in Figs.~\ref{fig1} and \ref{fig2}(c,d).

Furthermore, chiral phonon condensation simultaneously causes a structural phase transition. With a phonon polarization vector $\bm{\epsilon}_{\bm{q}, l}\sim \bm e_x+il\bm e_y$, a frozen $(q,l)$ mode has spatial helicity $-l \cdot\mathrm{sgn}(q)$. Consequently, the CDW modulation $\rho(\bm{r})$ caused by the $(m,l)=(-1,1)$ pair and the resulting lattice displacement $\delta\bm{u}(\bm{R})$ are given by [Fig.~\ref{fig2}(c)],
\begin{align}
\rho_{\mathrm{RH}}(\mathbf{r}) &= \rho_A\cos(Qz-2\varphi),
\\
\delta\bm{u}_{\mathrm{LH}}(\bm{R})&=u_A[\cos(Q R_z)\bm{e}_{x}-\sin(Q R_z)\bm{e}_{y}],
\end{align}
while the $(m,l)=(1,-1)$ pair leads to [Fig.~\ref{fig2}(d)] 
\begin{align}
\rho_{\mathrm{LH}}(\mathbf{r}) &= \rho_A\cos(Qz+2\varphi),
\\
\delta\bm{u}_{\mathrm{RH}}(\bm{R})&=u_A[\cos(Q R_z)\bm{e}_{x}+\sin(Q R_z)\bm{e}_{y}].
\end{align}
Here, $\rho_A$ and $u_A$ are their amplitudes. The RH/LH for $\rho$ and $\delta\bm{u}$ denotes the spatial helicity of the CDW pattern and the frozen lattice displacement. Note that the handedness of all orders is flipped in RH parent crystals. These expressions describe not only the helical modulation character with a period common to $\rho(\mathbf{r})$ and $\delta\bm{u}(\bm{r})$, but rather the winding character with the integer $\Delta m$~\cite{AN2}. In this way, although the single-$Q$ plane wave modulation cannot be chiral in itself, this winding CDW exhibits a macroscopic chiral structure in real space. Strictly speaking, the charge density modulation $\rho(\bm{r})$ includes higher-harmonic components $\cos(Qz+(\Delta m+3p)\varphi)$ with $ p\in\mathbb{Z}$, arising from the modulo $n$ nature of the CAM. The dominant winding component is  determined by the low-energy electronic structure.

\textit{Formulation in achiral crystals}.---We next consider achiral crystals. Here we assume $\hat{C}_3$-symmetric crystals and nesting of the Fermi surface along the $\Gamma$–A line, as in the chiral case~\cite{SM2}. As an example, Fig.~\ref{fig3}(a) shows a monatomic trigonal lattice. In this setting, because CAM is well-defined for electrons and axial phonons by $\hat{C}_3$ symmetry, the formulation developed for the $\hat{S}_3$-symmetric systems and the Hamiltonian Eq.~\eqref{Hamiltonian} can be applied. However, in the present achiral crystals, the presence of mirror symmetry perpendicular to the $z$ axis and $\mathcal{T}$ symmetry enforces degeneracies between the $m=\pm1$ electronic bands and between the $l=\pm1$ axial phonon modes, e.g., $\omega_{q_z,-l}=\omega_{q_z,l}$.

Therefore, although the transverse PT is favored under the condition $g_{\mathrm{L}}/g_{\mathrm{T}}<\omega_{Q,\mathrm{L}}/\omega_{Q,\mathrm{T}}$  [Fig.~\ref{fig3}(c)], the corresponding transition temperatures coincide $T_{\mathrm{C},+1}=T_{\mathrm{C},-1}\equiv T_{\mathrm{C}}$ due to the degeneracy of the $l=\pm1$ modes. An equal-amplitude superposition of the two degenerate axial modes gives an achiral nematic CDW:
\begin{align}
    \rho(\mathbf{r}) &= \rho'_A\cos(Qz)\cos(2\varphi),
    \\
    \delta\bm{u}(\bm{R})&=u'_A \cos(Q R_z)\bm{e}_{x}.
\end{align}
This phase does not exhibit chiral structure.

\textit{Spontaneous mirror symmetry breaking}.--- Nevertheless, this degeneracy suggests that a winding CDW can be stabilized once spontaneous symmetry breaking occurs. To examine this possibility, we derive the Landau free energy from the Hamiltonian in Eq.~\eqref{Hamiltonian} in terms of the order parameters $\Phi_{\pm1}$ corresponding to the condensed axial phonons with $l=\pm 1$ at the wave vector $Q$. Up to quartic order, it is given by~\cite{SM2}
\begin{align}
    \mathcal{F} =&\mathcal{F}_0 + \alpha_+|\Phi_{+1}|^2 + \alpha_- |\Phi_{-1}|^2 
    \notag \\
    &+ \beta_1 (|\Phi_{+1}|^4 + |\Phi_{-1}|^4) + \beta_2 |\Phi_{+1}|^2 |\Phi_{-1}|^2 .
    \label{Free_energy}
\end{align}
Here, $\alpha_{\pm} \simeq \, \alpha (T-T_{\mathrm{C}})$
arise from the same bubble diagrams as in the chiral system and determine the transition temperature, while the coefficients $\beta_1$ and $\beta_2$ encode the interaction between different CAM channels of electron bands (see End Matter).

Minimizing the free energy \eqref{Free_energy}, we obtain two competing ordered states as shown in Fig.~\ref{fig3}(b). When only a single axial phonon mode condenses, $|\Phi_{+1}| \neq 0$ or $|\Phi_{-1}| \neq 0$, the winding CDW is realized. The corresponding minimum of the free energy is given by $\mathcal{F}_{\mathrm{min}}^{\mathrm{W}}= \mathcal{F}_{0} - \alpha^2 (T_{\mathrm{C}}-T)^2/(4\beta_1)$. On the other hand, when both axial phonons condense with equal amplitude, $|\Phi_{+1}| = |\Phi_{-1}| \neq 0$, the achiral nematic CDW is realized. In this case, the minimum free energy is
$\mathcal{F}_{\mathrm{min}}^{\mathrm{Q}}= \mathcal{F}_{0} - \alpha^2 (T_{\mathrm{C}}-T)^2/(2\beta_1 + \beta_2)$. Therefore, the relative stability of these two phases is determined by the ratio of the quartic coefficients $\beta_1$ and $\beta_2$. The winding CDW is favored when $\beta_2 > 2\beta_1$, whereas the nematic CDW is stabilized for $\beta_2 < 2\beta_1$ [Fig.~\ref{fig3}(c)].

To examine the relative stability of these phases controlled by the ratio $\beta_2/\beta_1$, we evaluate this ratio from the diagrammatic expansion. We find that $\beta_2/\beta_1$ can exceed $2$ in a realistic parameter regime, indicating that the winding CDW is stabilized. This occurs when the $m=0$ electronic band lies close to the $m=\pm1$ bands or crosses the Fermi level, which enhances the inter-channel processes contributing to $\beta_2$.
As a representative example, we consider a minimal one-dimensional band structure given by $\xi_{k_z,0}=U_0-t_0 \cos k_z-\mu$ and $\xi_{k_z,\pm1}=U_{1}-t_{1}\cos k_z-\mu$. Fixing all other parameters, we vary $U_0$ to control the relative energy of the $m=0$ band. As shown in Fig.~\ref{fig3}(d), the resulting ratio $\beta_2/\beta_1$ as a function of $U_0$ demonstrates that $\beta_2/\beta_1$ exceeds $2$ in a finite parameter range. Figure~\ref{fig3}(e) displays the electronic bands for $U_0-\mu=-25$, where the condition $\beta_2/\beta_1>2$ is satisfied. In this regime, the proximity of the $m=0$ band to the Fermi level significantly enhances $\beta_2$ relative to $\beta_1$. Generally, $\beta_2/\beta_1$ is enhanced when a Fermi surface of the $m=0$ band is present.

These results demonstrate that multiband effects involving the $m=0$ channel play a crucial role in stabilizing the winding CDW. When the winding CDW phase is stabilized, the lattice displacement is dominated by a single axial phonon mode, leading to a helical lattice displacement. As a result, the crystal structure undergoes an achiral-to-chiral transition.

\textit{Discussion}.---We emphasize that our results are not restricted to the conventional $2k_{\mathrm F}$ instability. The essential ingredient is the CAM difference between electronic states connected by a nesting vector $\bm Q$ on the Fermi surface. Moreover, our formulation based on the eigenvalues of $\hat{C}_n$ or $\hat{S}_n$ can be applied to microscopic orbitals (see End Matter). In this sense, our framework is closely related to orbital-selective CDWs and may offer a unified and transparent perspective on such phenomena~\cite{Que2025NatCommun, Silvera-Vega2026CommunPhys, Antonelli2022npjQuantumMater, Xu2023npjQuantumMater, Singh2025NatPhys, McMahon2020SciAdv}. End Matter discusses SrAl$_4$-type compounds as candidate materials satisfying the microscopic prerequisites for our mechanism.

Experimentally, transverse phonon softening can be directly probed by using inelastic X-ray or neutron scattering~\cite{Hippert2005, Ament2011RevModPhys, Tingting2025arxiv}. To fully characterize the CDW state, it is pivotal to identify not only its chirality but also the winding character of the CDW. For example,
their character can be probed by coherent X-ray imaging techniques, such as Bragg coherent diffraction imaging or ptychography, as well as by phase-sensitive electron microscopy methods, including 4D-STEM and electron holography~\cite{Robinson2009NatMater, Colin2019Microscopy, Pfeiffer2018NatPhoto, Lichte2008ProgPhys}. In addition, nonlinear optical responses may provide complementary symmetry-sensitive fingerprints of the winding CDW state~\cite{Yang2025NatComm}.

Our work demonstrates a direct connection between structural chirality and the phase topology of electronic order. Furthermore, we provide insight into how chiral structures spontaneously emerge from the underlying electronic degrees of freedom~\cite{Budin2010AnnualRevBiophys,Michaeli2016ChemSocRev}. For applications, the winding CDW, characterized by a vortex-like phase structure analogous to vortex beams, opens a route toward novel functionalities in quantum electronics.

\textit{Acknowledgments}.---This work was supported by JSPS KAKENHI (Grant Numbers JP25KJ1482, JP22H04933, JP23K17353, JP23K22452, JP24K21530, JP24H00007, JP25H01249, JP26H02016).

\bibliography{WCDW}

\begin{thebibliography}{81}
\bibitem [{AN2()}]{AN2}%
  \BibitemOpen
  \href@noop {} {}\bibinfo {note} {By analogy with quantum vortices, the
  winding number is defined for the CDW order $\rho(\mathbf{r})=
  \rho(\mathbf{r}_{\perp},z) e^{i\Delta m \varphi} + \text{c.c.}$ as $\nu_{\rm
  C} = (2\pi)^{-1} \oint_C d\phi \, \partial_\phi \arg
  \rho(\mathbf{r}_{\perp},z) e^{i\Delta m \varphi}=\Delta m$, provided that
  $\rho(\mathbf{r}_{\perp},z)\neq 0$ on the loop C encircling the rotation or
  screw axis.}\BibitemShut {Stop}%
\end{thebibliography}


\onecolumngrid\begin{center}\textbf{\large{End Matter}}\end{center}

\begin{figure*}[t!]
    \centering
    \includegraphics[width=0.9\linewidth]{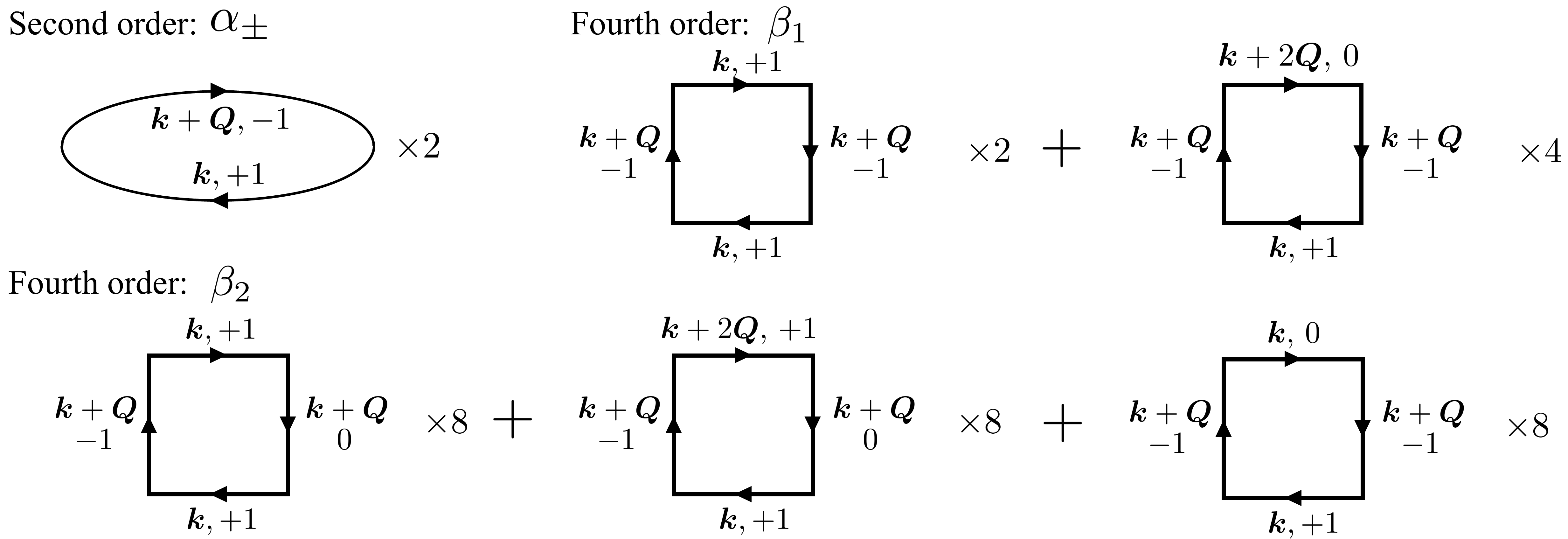}
    \caption{Feynman diagrams corresponding to each coefficient of the Landau free energy in Eq.~\eqref{Free_energy}.}
    \label{Fig4}
\end{figure*}

\twocolumngrid
\textit{Landau free energy}---By converting the Hamiltonian in Eq.~\eqref{Hamiltonian} into a path integral form and then applying the Hubbard-Stratonovich transformation, we can derive the free energy. In this process, expressions for the coefficients in Eq.~\eqref{Free_energy} are obtained. Let $\mathscr{G}^{(0)}(\bm{k},m,i\epsilon_{\ell})=1/(i\epsilon_{\ell}-\xi_{\bm{k},m})$ denote the Green’s function for the electron with fermionic Matsubara frequencies $i\epsilon_{\ell}$. In $\hat{C}_3$-symmetric crystals, the coefficients of the quadratic and quartic terms are given by the Feynman diagrams in Fig.~\ref{Fig4}, and the expressions are obtained as follows.
\begin{widetext}
\begin{align}
    \alpha_{+}&=\alpha_{-}=\hbar\omega_{\bm{Q},\pm1}+\dfrac{2g_{\mathrm{T}}^2}{N_{\mathrm{i}}\beta} \sum_{\bm{k}, i\epsilon_{\ell}}  \mathscr{G}^{(0)}(\bm{k}, +1, i\epsilon_{\ell}) \mathscr{G}^{(0)}(\bm{k}+\bm{Q}, -1, i\epsilon_{\ell}), 
    \\
    \beta_1 &=\dfrac{g_{\mathrm{T}}^4}{4}\dfrac{k_{\mathrm{B}}T}{N_{\mathrm{i}}} \sum_{\bm{k}, i\epsilon_{\ell}} \left[ 2 \mathscr{G}^{(0)}(\bm{k}, +1,i\epsilon_{\ell})^2 \mathscr{G}^{(0)}(\bm{k}+\bm{Q}, -1,i\epsilon_{\ell})^2 \right.
    \notag \\
    &\qquad\qquad\quad\;\; \left. + 4 \mathscr{G}^{(0)}(\bm{k}, +1,i\epsilon_{\ell}) \mathscr{G}^{(0)}(\bm{k}+\bm{Q}, -1,i\epsilon_{\ell})^2 \mathscr{G}^{(0)}(\bm{k}+2\bm{Q}, 0,i\epsilon_{\ell}) \right], 
    \label{eq_beta1}
    \\
    \beta_2 &=\dfrac{g_{\mathrm{T}}^4}{4}\dfrac{k_{\mathrm{B}}T}{N_{\mathrm{i}}}\sum_{\bm{k}, i\epsilon_{\ell}} \left[ 8 \mathscr{G}^{(0)}(\bm{k}, +1,i\epsilon_{\ell})^2 \mathscr{G}^{(0)}(\bm{k}+\bm{Q}, -1,i\epsilon_{\ell}) \mathscr{G}^{(0)}(\bm{k}+\bm{Q}, 0,i\epsilon_{\ell}) \right.
    \notag \\
    &\qquad\qquad\quad\;\; + 8 \mathscr{G}^{(0)}(\bm{k}, +1,i\epsilon_{\ell}) \mathscr{G}^{(0)}(\bm{k}+\bm{Q}, -1,i\epsilon_{\ell}) \mathscr{G}^{(0)}(\bm{k}+2\bm{Q}, +1,i\epsilon_{\ell}) \mathscr{G}^{(0)}(\bm{k}+\bm{Q}, 0,i\epsilon_{\ell}) 
    \notag \\
    &\qquad\qquad\quad\;\; \left. + 8 \mathscr{G}^{(0)}(\bm{k}, +1,i\epsilon_{\ell}) \mathscr{G}^{(0)}(\bm{k}+\bm{Q}, -1,i\epsilon_{\ell})^2 \mathscr{G}^{(0)}(\bm{k}, 0,i\epsilon_{\ell})  \right].
    \label{eq_beta2}
\end{align}
\end{widetext}
Equation~\eqref{eq_beta1} includes, in its first term, a contribution that passes only through the electronic bands $m=\pm1$ associated with nesting. On the other hand, Eq.~\eqref{eq_beta2} is a higher-order contribution in which, for all diagrams, the electron propagator passes through the band with CAM $m = 0$. From this perspective, 
it is expected that the relative position of the $m=0$ electron band is crucial to satisfy $\beta_2 > 2\beta_1$, a condition stabilizing the winding CDW in achiral crystals.

\textit{Perspective on CDWs in EuAl$_4$ and SrAl$_4$}---EuAl$_4$ and SrAl$_4$ provide possible materials contexts for the mechanism proposed in this work. These compounds exhibit a transverse Peierls transition~\cite{Yang2025NatComm,Wang2024CommPhys,Sitaram2024PRR,Haoyang2024PhysRevMater,Korshunov2024PhysRevB}. Although the nature of the CDW order remains under debate, a chiral CDW state is among the proposed candidates~\cite{Yang2025NatComm}. Their crystal structure is three-dimensional and belongs to the achiral space group $I4/mmm$.
However, the electronic states near the high-symmetry line play a dominant role in the CDW formation, allowing a natural extension of our framework to these materials. In the presence of spin-orbit coupling, including the electron spin naturally generalizes the CAM selection rule to
\begin{align}
    j_z+l=j_z' \quad \mod 4,
\end{align}
where $j_z$ and $j_z'$ denote the half-integer CAM of the initial and final electronic states, and $l$ is the phonon CAM. This relation determines which phonon mode can couple efficiently to the electronic states.

For acoustic phonons, the allowed CAM is restricted to $l=0,\pm1$ for any $\hat{C}_n$-symmetric crystal~\cite{AN1}. This follows from the fact that acoustic phonons are Goldstone modes whose $q\to0$ limit continuously connects to rigid real-space motions: the longitudinal mode carries $l=0$, while the two transverse modes carry $l=\pm1$. Therefore, irrespective of the rotational symmetry of the lattice, a soft transverse acoustic phonon naturally supplies the CAM $l=\pm1$ required to connect electronic states with a difference in the CAM $\Delta j_z = j'_z - j_z =\pm1$.

In EuAl$_4$ and SrAl$_4$, the experimentally observed nesting vector is $\bm{Q}=(0,0,Q)$, which preserves the in-plane wave vector $k_\perp$~\cite{Haoyang2024PhysRevMater, Korshunov2024PhysRevB, Kobata2016JPSJ}. Consequently, the nested initial and final states remain in the same rotational orbit around the $k_z$ axis, so that the CAM conservation on the $\Gamma$-Z line can be extended to the relevant states through rotational phase matching of the Bloch wave functions, as shown in the Supplementary Materials~\cite{SM2}. Nesting occurs predominantly between the $\mathrm{LD}_6$ ($j_z=\pm1/2$) and $\mathrm{LD}_7$ ($j_z=\pm3/2$) bands, whose orbital characters are mainly Al-$p_z$ and Al-$p_{\pm}$, respectively~\cite{Wang2024CommPhys}. Assuming spin-conserving electron-phonon scattering, we see that transverse acoustic phonons with $l=\pm1$ are compatible with interband mixing between these nested states, consistent with the orbital characters of the $\mathrm{LD}_6$ and $\mathrm{LD}_7$ bands. As a result, the transverse phonon modes selectively couple the nested $\mathrm{LD}_6$ and $\mathrm{LD}_7$ bands and enhance the EPC in this channel~\cite{Wang2024CommPhys, Sitaram2024PRR}.

Although the nested Fermi surface is also located away from the $\Gamma$-Z line, the corresponding Bloch states continuously inherit the rotational phase factor of the $\mathrm{LD}_6$ and $\mathrm{LD}_7$ bands. Consequently, the same phase matching condition remains effective for the finite $k_{\perp}$ nesting states, which favors interband coherence between electronic states with $\Delta j_z=\pm1$ (see Supplementary Note 2). The resulting interband coherence acquires an azimuthal phase factor $e^{\pm i(1+4\ell)\varphi}$, leading to chiral CDWs of the form
\begin{align}
\rho(\bm{r})\sim \cos\!\left(Qz\pm(1+4\ell)\varphi\right),
\end{align}
or, when the two chiral CDWs are superposed,
\begin{align}
\rho(\bm{r})\sim \cos(Qz)\cos\!\left((1+4\ell)\varphi\right),
\end{align}
where $\ell \in \mathbb{Z}$ labels the higher-harmonic components. The former corresponds to the winding CDW state discussed in the main text, while the latter corresponds to an achiral nematic CDW state. These results show that the nesting geometry and orbital character of the electronic states in EuAl$_4$ and SrAl$_4$ naturally satisfy the microscopic requirements for the emergence of the winding CDW due to the mechanism proposed in this work.

\newpage

\ifarXiv
  

\section*{Supplementary material}

# Supplementary Note 1

### Pseudo angular momentum and crystal angular momentum

Before proceeding, we clarify the terminology used in this Letter, because the definition of pseudo angular momentum (PAM) varies across the literature. We use PAM to refer to the eigenvalue of $\hat{C}_n$ or $\hat{S}_n$ symmetry operations acting on a Bloch state, which is a well-defined quantum number along high-symmetry lines satisfying $\hat{C}_n\boldsymbol{k} = \boldsymbol{k} + \boldsymbol{G}$, with $a$ reciprocal lattice vector $\boldsymbol{G}$ (e.g., $\boldsymbol{G} = 0$ on Γ-A line) [1–6]. We consider crystals possessing either $\hat{C}_n$ or $\hat{S}_n = \hat{C}_n\hat{T}_{c/n}$, where $\hat{T}_{c/n}$ denotes translation by $c/n$ along the rotational axis (here taken as the $z$-axis).

For a phonon mode propagating along the $z$-axis with wavevector $\boldsymbol{q} = (0, 0, q_z)$, the Bloch eigenvector $\boldsymbol{u}_{\boldsymbol{q}}$ satisfies

$$\hat{S}_n\boldsymbol{u}_{\boldsymbol{q}} = e^{-2\pi i l_{\rm PAM}/n}\boldsymbol{u}_{\boldsymbol{q}}, \tag{S.1}$$

where the PAM eigenvalue contains both a rotational and a translational contribution: $l_{\rm PAM} = l_{\rm ph} + q_z c/2\pi$. In the case of $\hat{S}_n$ symmetry, the additional term $q_z c/2\pi$ arising from the translational part of $\hat{S}_n$ renders $l_{\rm PAM}$ a $q$-dependent fractional quantity. However, the purely rotational part $l_{\rm ph}$ takes quantized integer values modulo $n$ and coincides with the PAM in the $\hat{C}_n$-symmetric case [7, 8]. Throughout this Letter, we refer to this integer $l_{\rm ph}$ as the crystal angular momentum (CAM). For electronic Bloch states, an analogous decomposition applies, and the electronic CAM $m$ is defined as the corresponding purely rotational integer part of the electronic PAM. Here, we consistently treat spinless electron bands, for which the CAM takes integer values modulo $n$. The extension to the spinful case, which requires a double-group treatment and leads to half-integer PAM, is discussed in Ref. [9].

When the interaction Hamiltonian respects the same $\hat{C}_n$ or $\hat{S}_n$ symmetry, and the scattering wavevector is directed along the rotational axis, the total PAM is conserved in each scattering process. Moreover, the translational contributions $q_z c/2\pi$ and $k_z c/2\pi$ appearing in the phononic and electronic PAM cancel each other as a consequence of crystal momentum conservation. As a result, the conservation law for the electron-phonon coupling reduces to that of the total CAM alone:

$$m + l = m' \mod n, \tag{S.2}$$

where $m$ and $m'$ are the CAM of the initial and final electronic states, respectively, and $l$ is the CAM of the phonon involved in the scattering process.

### Phonon displacement field and electron-phonon coupling

Since this Letter is concerned with the Peierls transition, we restrict our attention to acoustic phonon modes throughout. As Goldstone modes associated with the spontaneous breaking of continuous symmetry, acoustic phonons are gapless and, in the limit $q \to 0$, correspond to rigid real-space displacements of the lattice. Therefore, these modes are associated with both the real and pseudo angular momenta in the long-wavelength regime.

In the long-wavelength limit $q_z c \ll 1$, the phonon wavelength significantly exceeds the size of the unit cell, so the internal atomic structure within a unit cell is not resolved on the macroscopic observation scale. In general, the information of the CAM carried by a phonon eigenmode originates from the inter-site phase factor (intercell part) and the circular polarization character of the polarization vector (intracell part) [5, 6]. However, the former varies on the scale of the unit cell size, which becomes negligible compared to the phonon wavelength in the long-wavelength limit. Furthermore, for $\boldsymbol{q} = (0, 0, q_z)$ lying on the rotational axis, one has $\boldsymbol{G} = 0$ exactly in $\hat{C}_n\boldsymbol{q} = \boldsymbol{q} + \boldsymbol{G}$, and thus the intercell phase factor $e^{i\boldsymbol{G}\cdot\boldsymbol{R}_j}$ becomes unity for this wavevector. Consequently, in this limit, the inter-site phase is absorbed into a macroscopic effective polarization vector, and the CAM is carried solely by the circular polarization character of $\boldsymbol{\epsilon}_{\boldsymbol{q},l}$. Note that such a decomposition does not hold in general but serves merely as an aid to physical intuition. The displacement field therefore reduces to a simple Bloch sum parametrized only by the unit cell position, with the CAM label $l$ in one-to-one correspondence with the chirality of the polarization vector $\boldsymbol{\epsilon}_{\boldsymbol{q},l}$. Thus the phonon displacement field with wavevector $q_z$ and CAM $l$ can be written as

$$\boldsymbol{u}(\boldsymbol{R}) = \sum_{q_z,l}\sqrt{\frac{\hbar}{2N_{\rm i}M_{\rm i}\omega_{\boldsymbol{q},l}}}\boldsymbol{\epsilon}_{\boldsymbol{q},l}\left(\hat{b}_{\boldsymbol{q},l} + \hat{b}^{\dagger}_{-\boldsymbol{q},-l}\right)e^{iq_zR_z}, \tag{S.3}$$

where $M_{\rm i}$ is the ionic mass and $N_{\rm i}$ is the number of ions. The transverse modes with $l = \pm 1$ correspond to circularly polarized vibrations $\boldsymbol{\epsilon}_{q_z,\pm 1} \propto \boldsymbol{e}_x \pm i\boldsymbol{e}_y$. We distinguish the circular motion of a propagating chiral phonon from the spatial helicity of the

frozen displacement. With the convention $\boldsymbol{\epsilon}_{q_z,l} = (\mathbf{e}_x + il\mathbf{e}_y)/\sqrt{2}$ and the time dependence $e^{i(q_z z - \omega_q t)}$, the local circular motion has angular momentum $\hbar l$. After freezing, however, $\boldsymbol{u}_{Q,l} = \mathrm{Re}[\boldsymbol{\epsilon}_{Q,l} e^{iQz}]$ has spatial helicity $\mathrm{sgn}\, \partial_z \arg(u_x + iu_y) = -l\, \mathrm{sgn}(Q)$.

Meanwhile, the electron-phonon coupling (EPC) must obey the conservation law of CAM in Eq. (S.2) as long as the scattering direction is assumed to be parallel to the rotational axis. The EPC Hamiltonian is derived by expanding the ionic potential in powers of the displacement field and projecting onto the electronic Bloch states. Therefore, the full form of the EPC Hamiltonian is described as

$$\mathcal{H}_{\mathrm{ep}} = \frac{1}{\sqrt{N_{\mathrm{i}}}} \sum_{\boldsymbol{k},\boldsymbol{q}} \sum_{m,m',l} g_l(\boldsymbol{k},\boldsymbol{q}) \zeta_{\boldsymbol{q},l}\, \hat{c}^\dagger_{\boldsymbol{k}+\boldsymbol{q},m'} \hat{c}_{\boldsymbol{k},m} (\hat{b}_{\boldsymbol{q},l} + \hat{b}^\dagger_{-\boldsymbol{q},-l}) \delta(m+l, m'), \tag{S.4}$$

where $\zeta_{\boldsymbol{q},l} = \sqrt{\hbar/(2M_{\mathrm{i}}\omega_{\boldsymbol{q},l})}$ and $\delta(m+l,m')$ means the constraint $m + l = m' \mod n$. In general, it is difficult to determine the coupling constant $g_l(\boldsymbol{k},\boldsymbol{q})$. Instead, we follow the long-wavelength scheme described above and use the simple expression up to zeroth order in $k$ and first order in $q$, $g_l(\boldsymbol{k},\boldsymbol{q}) \simeq iqg_l$ [10]. Here, the $l = 0$ phonon can be regarded as a longitudinal oscillation, while $l = \pm 1$ as transverse oscillations. Therefore, the EPC Hamiltonian naturally decomposes into longitudinal and transverse sectors in the long-wavelength regime:

$$\mathcal{H}^{(\mathrm{L})}_{\mathrm{ep}} = \frac{1}{\sqrt{N_{\mathrm{i}}}} \sum_{\boldsymbol{k},\boldsymbol{q},m} iq\, g_{\mathrm{L}} \zeta_{\boldsymbol{q},0}\, \hat{c}^\dagger_{\boldsymbol{k}+\boldsymbol{q},m} \hat{c}_{\boldsymbol{k},m} \left( \hat{b}_{\boldsymbol{q},0} + \hat{b}^\dagger_{-\boldsymbol{q},0} \right), \tag{S.5}$$

$$\mathcal{H}^{(\mathrm{T})}_{\mathrm{ep}} = \frac{1}{\sqrt{N_{\mathrm{i}}}} \sum_{\boldsymbol{k},\boldsymbol{q},m} \sum_{\lambda=\pm 1} iq\, g_{\mathrm{T}} \zeta_{\boldsymbol{q},\lambda}\, \hat{c}^\dagger_{\boldsymbol{k}+\boldsymbol{q},m+\lambda} \hat{c}_{\boldsymbol{k},m} \left( \hat{b}_{\boldsymbol{q},\lambda} + \hat{b}^\dagger_{-\boldsymbol{q},-\lambda} \right), \tag{S.6}$$

where $g_{\mathrm{T}} \equiv g_\pm$ ($g_{\mathrm{L}} \equiv g_0$) is the coupling constant for the transverse (longitudinal) mode. Strictly speaking, $g_+$ and $g_-$ do not coincide in chiral crystals; however, when considering Peierls instability qualitatively, it is reasonable to assume that they are identical. This is because the inequivalence in the Peierls instability between the two chiral phonon channels is expected to be further enhanced when $g_+ \neq g_-$.

## Peierls instability in chiral crystals

The renormalized phonon frequency for each mode can be obtained within the Random Phase Approximation (RPA), as represented by the Feynman diagram in Fig. S1. The Hamiltonian employed in this RPA calculation is given by $\mathcal{H} = \mathcal{H}_{\mathrm{el}} + \mathcal{H}_{\mathrm{ph}} + \mathcal{H}_{\mathrm{ep}}$, where the single-particle electron and phonon Hamiltonians are $\mathcal{H}_{\mathrm{el}} = \sum_{\boldsymbol{k},m} \xi_{\boldsymbol{k},m} \hat{c}^\dagger_{\boldsymbol{k},m} \hat{c}_{\boldsymbol{k},m}$ and $\mathcal{H}_{\mathrm{ph}} = \sum_{\boldsymbol{q},l} \hbar\omega_{\boldsymbol{q},l} \hat{b}^\dagger_{\boldsymbol{q},l} \hat{b}_{\boldsymbol{q},l}$, respectively, and $\mathcal{H}_{\mathrm{ep}}$ is the EPC Hamiltonian derived in the previous section. The phonon self-energy for each mode with CAM $l$ is given by the electron-hole bubble diagram:

$$\Pi_l(\boldsymbol{q}, i\omega_{\ell'}) = \frac{k_{\mathrm{B}}T}{N_{\mathrm{i}}} \sum_{\boldsymbol{k},m,i\epsilon_\ell} |iqg_l\zeta_{\boldsymbol{q},l}|^2 \mathscr{G}^{(0)}(\boldsymbol{k}+\boldsymbol{q}, m+l, i\epsilon_\ell + i\omega_{\ell'}) \mathscr{G}^{(0)}(\boldsymbol{k}, m, i\epsilon_\ell), \tag{S.7}$$

where $\mathscr{G}^{(0)}(\boldsymbol{k}, m, i\epsilon_\ell) = (i\epsilon_\ell - \xi_{\boldsymbol{k},m})^{-1}$ is the bare electron Green's function. Performing the Matsubara frequency summation, we obtain

$$\frac{k_{\mathrm{B}}T}{N_{\mathrm{i}}} \sum_{\boldsymbol{k},m,i\epsilon_\ell} \mathscr{G}^{(0)}(\boldsymbol{k}+\boldsymbol{q}, m+l, i\epsilon_\ell + i\omega_{\ell'}) \mathscr{G}^{(0)}(\boldsymbol{k}, m, i\epsilon_\ell) = \frac{1}{N_{\mathrm{i}}} \sum_{\boldsymbol{k},m} \frac{f(\varepsilon_{\boldsymbol{k},m}) - f(\varepsilon_{\boldsymbol{k}+\boldsymbol{q},m+l})}{i\omega_{\ell'} + \varepsilon_{\boldsymbol{k},m} - \varepsilon_{\boldsymbol{k}+\boldsymbol{q},m+l}}, \tag{S.8}$$

where $f(\varepsilon)$ is the Fermi distribution function. Upon analytic continuation $i\omega_{\ell'} \to \hbar\omega + i\delta$, we define the Lindhard function for each mode $l$:

$$\mathcal{L}_l(\boldsymbol{q}, \omega) \equiv -\frac{1}{N_{\mathrm{i}}} \sum_{\boldsymbol{k},m} \frac{f(\varepsilon_{\boldsymbol{k},m}) - f(\varepsilon_{\boldsymbol{k}+\boldsymbol{q},m+l})}{\hbar\omega + i\delta + \varepsilon_{\boldsymbol{k},m} - \varepsilon_{\boldsymbol{k}+\boldsymbol{q},m+l}}. \tag{S.9}$$

The renormalized phonon frequency for mode $l$ then takes the form

$$(\hbar\tilde{\omega}_{q_z,l})^2 = (\hbar\omega_{q_z,l})^2 \left( 1 - \frac{q_z^2 g_l^2}{M_{\mathrm{i}} \omega^2_{q_z,l}} \mathcal{L}_l(q_z, \tilde{\omega}_{q_z,l}) \right). \tag{S.10}$$

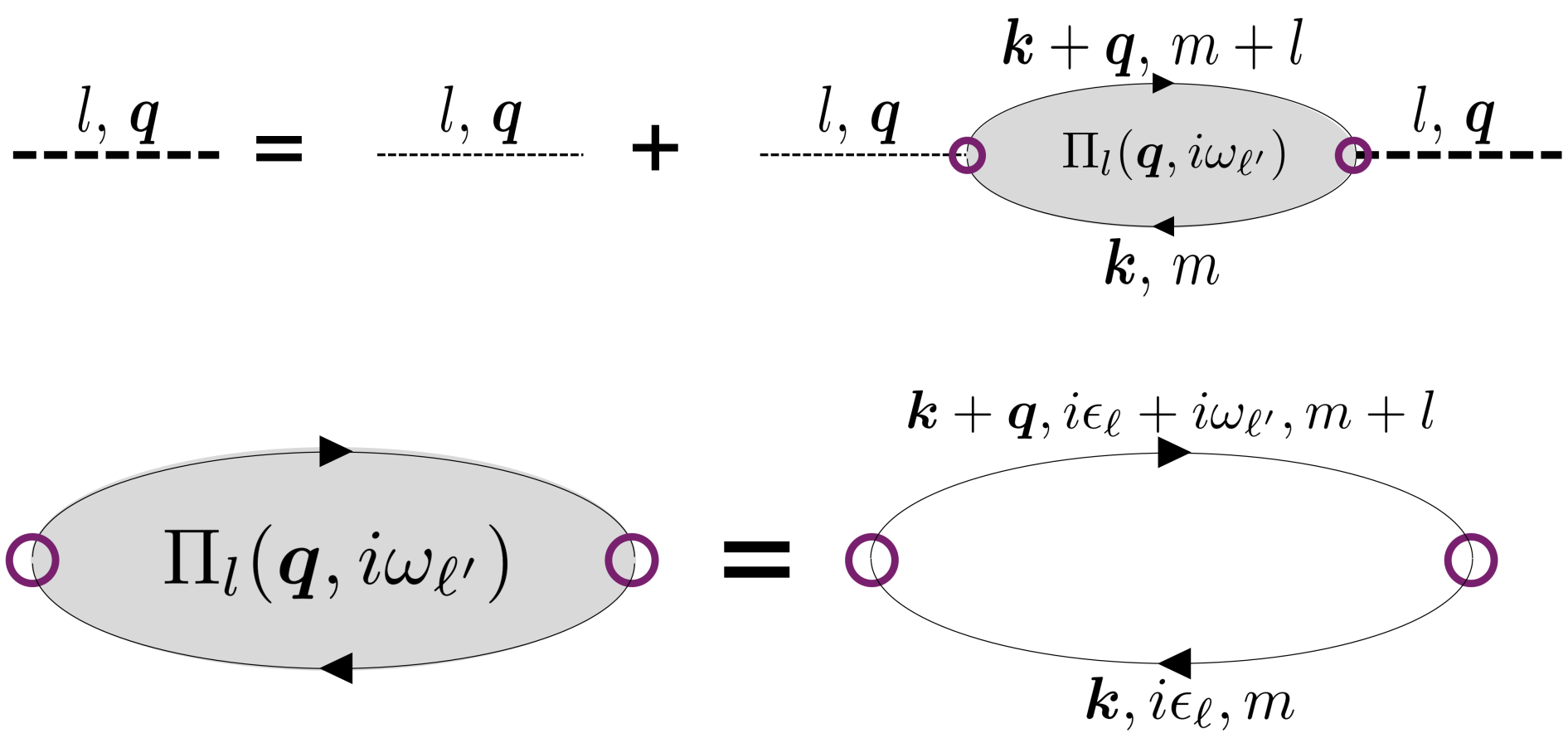


Figure S1. Feynman diagrams for the renormalized phonon frequencies. (Top) Dyson equation for the phonon propagator with CAM $l$ and wavevector $\boldsymbol{q}$: the renormalized propagator (thick dashed line) is expressed as the bare propagator (thin dashed line) dressed by the phonon self-energy $\Pi_l(\boldsymbol{q}, i\omega_{\ell'})$. (Bottom) Explicit diagrammatic representation of the self-energy $\Pi_l(\boldsymbol{q}, i\omega_{\ell'})$ within the RPA, given by the electron-hole bubble: solid lines with arrows denote the bare electron Green's function $\mathscr{G}^{(0)}(\boldsymbol{k}, m, i\epsilon_\ell)$, and purple circles denote the electron-phonon coupling vertex $iqg_l\zeta_{\boldsymbol{q},l}$.

For simplicity, we work in the static limit $\omega \to 0$, where the Lindhard function reduces to

$$\mathcal{L}_l(q_z, 0) = -\frac{1}{N_\mathrm{i}} \sum_{k_z,m} \frac{f(\varepsilon_{k_z,m}) - f(\varepsilon_{k_z+q_z,m+l})}{\varepsilon_{k_z,m} - \varepsilon_{k_z+q_z,m+l} + i\delta}. \tag{S.11}$$

The instability for each mode $l$ is thus governed by the Lindhard function $\mathcal{L}_l(q_z, 0)$, which is enhanced when electronic states connected by $\boldsymbol{q}$ satisfy the CAM conservation law. In chiral crystals, neither the electronic bands nor the phonon branches are degenerate between sectors with opposite CAM, so the Lindhard function—and hence the degree of phonon softening—is in general inequivalent between different modes $l$. Consequently, the instability is most strongly developed for the phonon mode whose CAM is matched to the Fermi surface via the conservation law, Eq. (S.2). Note that at this stage of the analysis, we only use the information of the electronic bands, the phonon dispersion, and the EPC Hamiltonian derived in the previous section; in particular, the explicit form of electronic wave functions has not yet been specified, and the only approximation invoked is the long-wavelength assumption of phonon inherent in the EPC.

To perform explicit calculations, we specialize to the $\hat{S}_3$-symmetric chiral crystal and adopt the electronic band dispersion

$$\xi_{k_z,m} = -t\cos\left(\frac{k_z c}{3} + \frac{2m\pi}{3}\right) - \mu, \tag{S.12}$$

which gives bands with CAM $m = 0, \pm 1$ that are intrinsically split in chiral crystals [11]. Since the electronic band structure alone already lifts the equivalence between different CAM sectors, the phonon softening is rendered inequivalent between modes regardless of whether the bare phonon dispersions themselves are split. In actual chiral crystals, while the $l = \pm 1$ transverse acoustic phonons are degenerate at the level of the sound velocity, their chiral splitting in $\omega_{q_z,l}$ appears in the order $O(q_z^2)$, which would further enhance the inequivalence of the two transverse Peierls instabilities [12]. For simplicity, we therefore adopt a phonon dispersion without chiral splitting: $\omega_{q_z,0} = v_\mathrm{L} q_z$ and $\omega_{q_z,\pm 1} = v_\mathrm{T} q_z$, with $v_\mathrm{L} = 1$ and $v_\mathrm{T} = 0.35$. The remaining parameters are set to $t = 2$, $c = 1$, $g_\mathrm{T} = 0.25$, $g_\mathrm{L} = 0.5$, and $M_\mathrm{i} = 1$. The resulting renormalized phonon dispersions at the respective Peierls transition temperatures are shown in Fig. 2(b) of the main text. We consider $\mu = -1.9$, $0.5$, $1.3$ illustrated by X, Y and Z in Fig. 2(b), respectively, where different types of Fermi surface instability are predicted.

### Expression of CDW orders

In this section, we show the explicit forms of the winding CDW and the accompanying lattice displacement arising from the transverse Peierls transition driven by the $l = \pm 1$ chiral phonon condensation. As a concrete example, we consider the case in which the instability driven by $\mathcal{L}_{l=+1}(q_z, 0)$ leads to the condensation of chiral phonon pairs with $(l = +1, q_z = Q)$ and $(l = -1, q_z = -Q)$. This corresponds to case Y in Fig. 2(b) of the main text, where the electronic states near $-k_\mathrm{F}$ $(m = +1)$ and $k_\mathrm{F}$ $(m = -1)$ are coupled by the nesting vector $Q$.

We first derive the lattice displacement induced by the chiral phonon condensation. Since only the phonon field operators $\zeta_{\boldsymbol{q},l}(\hat{b}_{\boldsymbol{q},l} + \hat{b}^{\dagger}_{-\boldsymbol{q},-l})$ for $(l = +1, q_z = Q)$ and $(l = -1, q_z = -Q)$ acquire finite expectation values, Eq. (S.3) reduces to

$$\begin{aligned}\delta\boldsymbol{u}(\boldsymbol{R}) &\equiv \langle \boldsymbol{u}(\boldsymbol{R})\rangle \\ &= \frac{1}{\sqrt{N_{\mathrm{i}}}}\boldsymbol{\epsilon}_{Q,+1}\langle \zeta_{Q,+1}\hat{b}_{Q,+1} + \zeta_{-Q,-1}\hat{b}^{\dagger}_{-Q,-1}\rangle e^{iQR_z} + \text{c.c.} \\ &= \frac{|\Delta|}{\sqrt{2}Qg_{\mathrm{T}}}\begin{bmatrix}\cos(QR_z+\theta_0)\\ -\sin(QR_z+\theta_0)\\ 0\end{bmatrix}.\end{aligned} \tag{S.13}$$

Here, $|\Delta|$ is the order parameter corresponding to the Peierls gap, and $\theta_0$ denotes the initial phase offset of the modulation relative to the phase of the CDW order, which is omitted in the main text for simplicity. Equation (S.13) describes the static distortion due to freezing of the $(q_z, l) = (Q, +1)$ phonon mode. Since the frozen lattice helicity is $-l\,\mathrm{sgn}(q_z)$, this distortion forms a LH lattice helix. It is associated with the RH winding CDW in the $(m, l) = (-1, +1)$ channel.

Next, we derive the real-space charge density profile of the winding CDW. The charge density operator is expressed in terms of the field operator $\hat{\Psi}(\boldsymbol{r}) = N_{\mathrm{i}}^{-1/2}\sum_{m,\boldsymbol{k}}\psi_{m,\boldsymbol{k}}(\boldsymbol{r})\hat{c}_{m,\boldsymbol{k}}$ as $\hat{\rho}(\boldsymbol{r}) = \hat{\Psi}^{\dagger}(\boldsymbol{r})\hat{\Psi}(\boldsymbol{r})$, which in momentum space takes the form

$$\hat{\rho}(\boldsymbol{r}) = \frac{1}{N_{\mathrm{i}}}\sum_{\boldsymbol{k},\boldsymbol{k}'}\sum_{m,m'} c^{\dagger}_{\boldsymbol{k}',m'}c_{\boldsymbol{k},m}\,\psi^{*}_{\boldsymbol{k}',m'}(\boldsymbol{r})\,\psi_{\boldsymbol{k},m}(\boldsymbol{r}). \tag{S.14}$$

Here, the electronic Bloch states along the $z$-axis in the $\hat{S}_n$-symmetric chiral crystal are labeled by the crystal wavevector $\boldsymbol{k}$ and the CAM $m$, and take the general form

$$\psi_{\boldsymbol{k},m}(\boldsymbol{r}) = e^{ik_z z}e^{im\varphi}u_{\boldsymbol{k},m}(r_{\perp},\varphi,z), \tag{S.15}$$

where $\varphi$ is the azimuthal angle, $r_{\perp}$ is the radial distance from the $z$-axis, and $u_{\boldsymbol{k},m}$ is the periodic part of the Bloch function satisfying

$$u_{\boldsymbol{k},m}(r_{\perp},\varphi,z) = u_{\boldsymbol{k},m}(r_{\perp},\varphi - \frac{2\pi}{n}, z - \frac{c}{n}). \tag{S.16}$$

The phase factor $e^{im\varphi}$ explicitly encodes the CAM $m$ of the electronic state.

Since the chiral phonon condensation that drives the lattice displacement also couples the electronic states $(m = +1, -k_{\mathrm{F}})$ and $(m = -1, k_{\mathrm{F}})$, only the anomalous expectation value $\langle c^{\dagger}_{k_{\mathrm{F}}+k,-1}c_{-k_{\mathrm{F}}+k,+1}\rangle$ acquires a macroscopically finite value. Retaining only this contribution and substituting the explicit Bloch wavefunctions Eq. (S.15), one obtains

$$\begin{aligned}\langle\hat{\rho}(\boldsymbol{r})\rangle &= \frac{1}{N_{\mathrm{i}}}\sum_{\boldsymbol{k}}\langle c^{\dagger}_{\boldsymbol{k}+\boldsymbol{Q},-1}c_{\boldsymbol{k},+1}\rangle\,\psi^{*}_{\boldsymbol{k}+\boldsymbol{Q},-1}(\boldsymbol{r})\,\psi_{\boldsymbol{k},+1}(\boldsymbol{r}) + \text{c.c.} \\ &= \frac{1}{N_{\mathrm{i}}}\sum_{\boldsymbol{k}} e^{-iQz}e^{i\varphi}u^{*}_{\boldsymbol{k}+\boldsymbol{Q},-1}(\boldsymbol{r})\,e^{i\varphi}u_{\boldsymbol{k},+1}(\boldsymbol{r})\,\langle c^{\dagger}_{\boldsymbol{k}+\boldsymbol{Q},-1}c_{\boldsymbol{k},+1}\rangle + \text{c.c.} \\ &= F(\boldsymbol{r})\,e^{-i(Qz-2\varphi)} + \text{c.c.},\end{aligned} \tag{S.17}$$

where the form factor is defined as

$$\begin{aligned}F(\boldsymbol{r}) =& \frac{1}{N_{\mathrm{i}}}\sum_{\boldsymbol{k}} u^{*}_{\boldsymbol{k}+\boldsymbol{Q},-1}(\boldsymbol{r})\,u_{\boldsymbol{k},+1}(\boldsymbol{r})\,\langle c^{\dagger}_{\boldsymbol{k}+\boldsymbol{Q},-1}c_{\boldsymbol{k},+1}\rangle \\ \simeq& \frac{\Delta}{N_{\mathrm{i}}}\sum_{\boldsymbol{k}} u^{*}_{\boldsymbol{k}+\boldsymbol{Q},-1}(\boldsymbol{r})\,u_{\boldsymbol{k},+1}(\boldsymbol{r}).\end{aligned} \tag{S.18}$$

Because $F(\boldsymbol{r})$ varies on the scale of the unit cell, which is much shorter than the modulation period $2\pi/Q$, it may be treated as approximately constant $\rho_{\mathrm{A}}$ on the scale of the CDW modulation.

It should be noted that the CAM $m$ is defined only modulo $n$, so that the periodic part of the Bloch function admits a Fourier expansion in the azimuthal angle of the form

$$u_{\boldsymbol{k},m}(\boldsymbol{r}) = \sum_{p\in\mathbb{Z}} f_{\boldsymbol{k},m,p}(r_{\perp},z)e^{i3p\varphi}.$$

Substituting this expansion into the expression for $\langle\hat{\rho}(\boldsymbol{r})\rangle$, one finds that the charge density can contain higher harmonic components,

$$\begin{aligned}\langle\hat{\rho}(\boldsymbol{r})\rangle =& \frac{1}{N_{\mathrm{i}}}\sum_{k}\sum_{p_1,p_2\in\mathbb{Z}} f^{*}_{k+Q,-1,p_2}(r_\perp,z) f_{k,+1,p_1}(r_\perp,z)\langle c^{\dagger}_{k+Q,-1}c_{k,+1}\rangle e^{-iQz}e^{i[2+3(p_1-p_2)]\varphi} + \text{c.c.}\\ =& \sum_{p\in\mathbb{Z}} F_p(r_\perp,z)e^{-iQz}e^{i(2+3p)\varphi} + \text{c.c.}\\ \simeq& \sum_{p\in\mathbb{Z}} A_p\cos\left(Qz-(2+3p)\varphi\right),\end{aligned} \tag{S.19}$$

where $F_p(r_\perp, z)$ is the form factor for each harmonic. Thus, the winding CDW generically consists of multiple angular harmonics $\cos(Qz + (\Delta m + 3p)\varphi)$ with $p \in \mathbb{Z}$, rather than a single sinusoidal component. The dominant harmonic, and hence the effective winding number, is expected to be determined by the orbital character of the electronic states near the Fermi surface that is essential for the Peierls instability.

In addition, the achiral nematic CDW can be realized in achiral crystals discussed in Supplementary Note 2, and its real-space structure follows straightforwardly from the same framework. When both $l = +1$ and $l = -1$ axial phonon channels condense simultaneously with equal amplitude, the total charge density and the lattice displacement are obtained by superposing the two contributions. The charge density becomes

$$\begin{aligned}\rho(\boldsymbol{r}) &= \rho_{\mathrm{RH}}(\boldsymbol{r}) + \rho_{\mathrm{LH}}(\boldsymbol{r})\\ &= 2\rho_A\cos(Qz)\cos(2\varphi)\\ &\equiv \rho'_A\cos(Qz)\cos(2\varphi),\end{aligned} \tag{S.20}$$

and the lattice displacement becomes

$$\begin{aligned}\delta\boldsymbol{u}(\boldsymbol{R}) &= \delta\boldsymbol{u}_{\mathrm{RH}}(\boldsymbol{R}) + \delta\boldsymbol{u}_{\mathrm{LH}}(\boldsymbol{R})\\ &= \frac{\sqrt{2}|\Delta|}{Qg_{\mathrm{T}}}\begin{bmatrix}\cos(QR_z+\theta_0)\\ 0\\ 0\end{bmatrix}\\ &\equiv u'_A\cos(QR_z+\theta_0)\boldsymbol{e}_x.\end{aligned} \tag{S.21}$$

The charge density thus takes the form of an achiral nematic modulation, while the helical components of the two lattice distortions cancel one another, leaving a transverse linearly polarized displacement.

# Supplementary Note 2

### Model for achiral crystals

We consider achiral crystals with $\hat{C}_3$ symmetry and assume nesting along the $\boldsymbol{k} = (0, 0, k_z)$ direction, e.g., the Γ–A and Γ–Z line. Although the crystal is no longer chiral, CAM remains a well-defined quantum number along this high-symmetry line, since the $\hat{C}_3$ rotation satisfies $\hat{C}_3\boldsymbol{k} = \boldsymbol{k}$. Moreover, the CAM conservation law in EPC, Eq. (S.2), holds for any $\hat{C}_n$-symmetric crystal regardless of whether a screw component is present. Therefore, the EPC Hamiltonian derived in Supplementary Note 1 applies without modification, and the full Hamiltonian retains the same form as

$$
\begin{aligned}
\mathcal{H} =& \mathcal{H}_{\rm el} + \mathcal{H}_{\rm ph} + \mathcal{H}_{\rm ep}^{(\rm L)} + \mathcal{H}_{\rm ep}^{(\rm T)}, \\
\mathcal{H}_{\rm el} =& \sum_{\boldsymbol{k},m} \xi_{\boldsymbol{k},m} \hat{c}^\dagger_{\boldsymbol{k},m} \hat{c}_{\boldsymbol{k},m}, \\
\mathcal{H}_{\rm ph} =& \sum_{\boldsymbol{q},l} \hbar\omega_{\boldsymbol{q},l} \hat{b}^\dagger_{\boldsymbol{q},l} \hat{b}_{\boldsymbol{q},l}, \\
\mathcal{H}_{\rm ep}^{(\rm L)} =& \frac{1}{\sqrt{N_{\rm i}}} \sum_{\boldsymbol{k},\boldsymbol{q},m} iqg_{\rm L}\zeta_{\boldsymbol{q},0}\, \hat{c}^\dagger_{\boldsymbol{k}+\boldsymbol{q},m} \hat{c}_{\boldsymbol{k},m} (\hat{b}_{\boldsymbol{q},0} + \hat{b}^\dagger_{-\boldsymbol{q},0}), \\
\mathcal{H}_{\rm ep}^{(\rm T)} =& \frac{1}{\sqrt{N_{\rm i}}} \sum_{\boldsymbol{k},\boldsymbol{q},m} \sum_{l=\pm 1} iqg_{\rm T}\zeta_{\boldsymbol{q},l}\, \hat{c}^\dagger_{\boldsymbol{k}+\boldsymbol{q},m+l} \hat{c}_{\boldsymbol{k},m} (\hat{b}_{\boldsymbol{q},l} + \hat{b}^\dagger_{-\boldsymbol{q},-l}).
\end{aligned}
$$

However, in contrast to the chiral system, the presence of mirror or inversion symmetry combined with time-reversal symmetry enforces degeneracies between the $m = \pm 1$ electronic bands and between the $l = \pm 1$ axial phonon modes. Here we adopt the following simple band dispersions along the Γ-A line:

$$
\xi_{k_z,0} = -t_0 \cos(k_z c) + U_0 - \mu, \qquad \xi_{k_z,\pm 1} = -t_1 \cos(k_z c) + U_1 - \mu. \tag{S.22}
$$

The analysis of the Peierls instability and the transition temperature for each phonon mode proceeds in the same way as in Supplementary Note 1. However, an important distinction arises in achiral crystals: even when the Fermi level lies on the $m = \pm 1$ bands, both the instability of $l = 0$ and $l = \pm 1$ phonon channels are symmetry-allowed, since the CAM conservation law does not by itself select a unique channel. The transverse Peierls transition is therefore realized only when the instability of the transverse phonon modes exceeds that of the longitudinal mode, which requires the condition

$$
\frac{g_{\rm L}}{g_{\rm T}} < \frac{\omega_{Q,\rm L}}{\omega_{Q,\rm T}}. \tag{S.23}
$$

Even assuming this condition is satisfied, the degeneracy between the $l = +1$ and $l = -1$ axial phonon modes indicates that the two channels have identical transition temperatures, $T_{\rm C,+1} = T_{\rm C,-1}$, within the RPA. Consequently, the microscopic calculation for the instability cannot select a definite handedness, and one must go beyond the quadratic level by constructing the Landau free energy to determine which ordered state is stabilized, as discussed in the following sections.

### Derivation of Landau free energy

While the transition temperature and the Peierls instability have been discussed based on the microscopic Hamiltonian so far, such an approach is not sufficient in achiral crystals. Therefore, we reformulate the problem within the path-integral framework and construct the Landau free energy. We first rewrite the Hamiltonian in terms of Grassmann variables and perform the Fourier transformation to momentum–frequency space. The fermionic and bosonic Matsubara frequencies are defined as $\epsilon_\ell = (2\ell + 1)\pi/\beta$ and $\Omega_\nu = 2\nu\pi/\beta$, respectively. In this section, we introduce the shorthand notation $k \equiv \{\boldsymbol{k}, i\omega_n\},\ q \equiv \{\boldsymbol{q}, i\Omega_\nu\}$. The creation and annihilation operators are replaced as $\hat{c}^\dagger, \hat{c} \to \bar{\psi}, \psi,\ \hat{b}^\dagger, \hat{b} \to \bar{\phi}, \phi$.

With these definitions, the partition function and the action are written as:

$$\mathcal{Z} = \int \mathcal{D}[\bar{\psi}, \psi, \bar{\phi}, \phi] \exp\left(-S_{\rm el}[\bar{\psi}, \psi] - S_{\rm ph}[\bar{\phi}, \phi] - S_{\rm ep}[\bar{\psi}, \psi, \bar{\phi}, \phi]\right), \tag{S.24}$$

$$S_{\rm el}[\bar{\psi}, \psi] = \sum_{k,m} \bar{\psi}_{k,m}(-i\epsilon_\ell + \varepsilon_{\boldsymbol{k},m} - \mu)\psi_{k,m}, \tag{S.25}$$

$$S_{\rm ph}[\bar{\phi}, \phi] = \sum_{q,\lambda} \bar{\phi}_{q,\lambda}(-i\Omega_\nu + \hbar\omega_{\boldsymbol{q},\lambda})\phi_{q,\lambda}, \tag{S.26}$$

$$S_{\rm ep}[\bar{\psi}, \psi, \bar{\phi}, \phi] = \frac{g_{\rm T}}{\sqrt{N_{\rm i}\beta}} \sum_{k,q,m,\lambda} \bar{\psi}_{k+q,m+\lambda}\left(\phi_{q,\lambda} + \bar{\phi}_{-q,-\lambda}\right)\psi_{k,m}. \tag{S.27}$$

We then perform the Gaussian integration over the fermionic fields $\bar{\psi}, \psi$, which yields

$$\mathcal{Z} = \int \mathcal{D}[\bar{\phi}, \phi] e^{-S_{\rm ph}[\bar{\phi},\phi]} \det \mathcal{M} \tag{S.28}$$

$$= \int \mathcal{D}[\bar{\phi}, \phi] \exp\left[-S_{\rm ph}[\bar{\phi}, \phi] + \operatorname{tr}\log \mathcal{M}[\bar{\phi}, \phi]\right], \tag{S.29}$$

where the matrix $\mathcal{M}$ is defined in $(\boldsymbol{k}, i\epsilon_\ell, m)$ space (combined indices of wave number, Matsubara frequency and CAM):

$$\mathcal{M}[\bar{\phi}, \phi]_{(k',m')(k,m)} = (-i\epsilon_\ell + \varepsilon_{\boldsymbol{k},m} - \mu)\delta_{k'k}\delta_{m'm} + \frac{g_{\rm T}}{\sqrt{N_{\rm i}\beta}} \sum_{q,\lambda} (\phi_{q,\lambda} + \bar{\phi}_{-q,-\lambda})\delta_{k',k+q}\delta_{m',m+\lambda}. \tag{S.30}$$

To analyze the phonon softening, we adopt the static approximation $\Omega_\nu = 0, \boldsymbol{q} = \boldsymbol{Q}$, and apply the saddle-point approximation by substituting

$$\phi_{q,\lambda} = \sqrt{N_{\rm i}\beta}\, \Phi_\lambda \delta_{\boldsymbol{q},\boldsymbol{Q}} \delta_{\Omega_\nu,0}. \tag{S.31}$$

The partition function is then approximated as

$$\mathcal{Z} \simeq \exp\left[-S_{\rm eff}[\bar{\Phi}, \Phi]\right] \tag{S.32}$$

$$= \exp\left[-\beta N_{\rm i} \sum_\lambda \hbar\omega_{\boldsymbol{Q},\lambda} |\Phi_\lambda|^2 + \operatorname{tr}\log \mathcal{M}[\bar{\Phi}, \Phi]\right]. \tag{S.33}$$

Accordingly, the free energy per ion is given by

$$\mathcal{F} = -\frac{1}{\beta N_{\rm i}} \log \mathcal{Z} = \sum_\lambda \hbar\omega_{\boldsymbol{Q},\lambda} |\Phi_\lambda|^2 - \frac{1}{\beta N_{\rm i}} \operatorname{tr}\log \mathcal{M}[\bar{\Phi}, \Phi], \tag{S.34}$$

where

$$\mathcal{M}[\bar{\Phi}, \Phi] = -\mathscr{G}^{(0)}(k)^{-1} + \mathcal{V}[\bar{\Phi}, \Phi] \tag{S.35}$$

$$= -\mathscr{G}^{(0)}(k)^{-1}\left(\mathbf{1} - \mathscr{G}^{(0)}(k)\mathcal{V}[\bar{\Phi}, \Phi]\right), \tag{S.36}$$

$$\mathcal{V}[\bar{\Phi}, \Phi]_{(k'm')(km)} = g_{\rm T} \sum_\lambda \left(\Phi_\lambda \delta_{\boldsymbol{k}',\boldsymbol{k}+\boldsymbol{Q}} + \bar{\Phi}_{-\lambda}\delta_{\boldsymbol{k}',\boldsymbol{k}-\boldsymbol{Q}}\right)\delta_{m',m+\lambda}\delta_{i\epsilon'_\ell, i\epsilon_\ell}. \tag{S.37}$$

Therefore, by expanding the matrix $\mathcal{M}$ in powers of $\mathcal{V}$, we obtain the series expansion of the free energy as

$$\mathcal{F} = \sum_\lambda \hbar\omega_{\boldsymbol{Q},\lambda}|\Phi_\lambda|^2 - \frac{1}{\beta N_{\rm i}} \operatorname{tr}\log(-\mathscr{G}^{(0)-1}) + \frac{1}{\beta N_{\rm i}} \sum_{j=1}^{\infty} \frac{1}{j} \operatorname{tr}(\mathscr{G}^{(0)}\mathcal{V})^j. \tag{S.38}$$

Let us consider the Landau free energy up to the fourth order of the order parameter of condensed axial phonons $\Phi_\lambda$. The free energy up to the fourth-order term becomes

$$\mathcal{F}[\bar{\Phi}, \Phi] = \mathcal{F}^{(0)} + \mathcal{F}^{(2)}[\bar{\Phi}, \Phi] + \mathcal{F}^{(4)}[\bar{\Phi}, \Phi]. \tag{S.39}$$

Here, $\mathcal{F}^{(0)}$ denotes the free energy of the non-interacting electrons:

$$\mathcal{F}^{(0)} = -\frac{1}{\beta N_{\rm i}} \mathrm{tr} \log(-\mathscr{G}^{(0)-1}). \tag{S.40}$$

In order to evaluate the second-order term $\mathcal{F}^{(2)}[\bar{\Phi}, \Phi]$, we compute

$$\mathrm{tr}(\mathscr{G}^{(0)}\mathcal{V})^2 = \sum_{k_1,m_1,\lambda_1} \sum_{k_2,m_2,\lambda_2} \mathscr{G}^{(0)}(k_1,m_1)\mathcal{V}_{(k_1,m_1,\lambda_1),(k_2,m_2,\lambda_2)}\mathscr{G}^{(0)}(k_2,m_2)\mathcal{V}_{(k_2,m_2,\lambda_2),(k_1,m_1,\lambda_1)}, \tag{S.41}$$

as follows. Focusing on the summation over the CAM index, we obtain

$$\sum_{\lambda_1,\lambda_2} \sum_{m_1,m_2} \delta_{m_1,m_2+\lambda_1}\delta_{m_2,m_1+\lambda_2} = \sum_{\lambda_1,\lambda_2} \sum_{m_1} \delta_{m_1,m_1+\lambda_1+\lambda_2}.$$

This implies that only contributions with $\lambda_2 = -\lambda_1$ survive in $\hat{C}_3$- (or $\hat{S}_3$-)symmetric systems. Consequently,

$$\begin{aligned}
\mathrm{tr}[(\mathscr{G}^{(0)}\mathcal{V})^2] =& \sum_{\lambda} g_{\rm T}^2 \sum_{\boldsymbol{k}_1,\boldsymbol{k}_2,m_1,m_2 i\epsilon_\ell} \mathscr{G}^{(0)}(\boldsymbol{k}_1, i\epsilon_\ell, m_1)\mathscr{G}^{(0)}(\boldsymbol{k}_2, i\epsilon_\ell, m_2) \\
&\times \left[\bar{\Phi}_{-\lambda}\Phi_\lambda \delta_{\boldsymbol{k}_1,\boldsymbol{k}_2+\boldsymbol{Q}}\delta_{\boldsymbol{k}_2,\boldsymbol{k}_1+\boldsymbol{Q}} + \bar{\Phi}_\lambda\bar{\Phi}_{-\lambda}\delta_{\boldsymbol{k}_1,\boldsymbol{k}_2-\boldsymbol{Q}}\delta_{\boldsymbol{k}_2,\boldsymbol{k}_1-\boldsymbol{Q}} + \bar{\Phi}_\lambda\Phi_\lambda\delta_{\boldsymbol{k}_1,\boldsymbol{k}_2-\boldsymbol{Q}}\delta_{\boldsymbol{k}_2,\boldsymbol{k}_1+\boldsymbol{Q}} + \bar{\Phi}_{-\lambda}\bar{\Phi}_{-\lambda}\delta_{\boldsymbol{k}_1,\boldsymbol{k}_2+\boldsymbol{Q}}\delta_{\boldsymbol{k}_2,\boldsymbol{k}_1-\boldsymbol{Q}}\right] \\
=& \sum_{\lambda} g_{\rm T}^2 \sum_{\boldsymbol{k}_1,\boldsymbol{k}_2,m,i\epsilon_\ell} \mathscr{G}^{(0)}(\boldsymbol{k}_1, m, i\epsilon_\ell)\mathscr{G}^{(0)}(\boldsymbol{k}_2, m+\lambda, i\epsilon_\ell)\left[|\Phi_\lambda|^2\delta_{\boldsymbol{k}_2,\boldsymbol{k}_1+\boldsymbol{Q}} + |\bar{\Phi}_{-\lambda}|^2\delta_{\boldsymbol{k}_1,\boldsymbol{k}_2+\boldsymbol{Q}}\right] \\
=& \sum_{\lambda} g_{\rm T}^2 \left(\sum_{\boldsymbol{k},m,i\epsilon_\ell} |\Phi_\lambda|^2 \mathscr{G}^{(0)}(\boldsymbol{k}, m, i\epsilon_\ell)\mathscr{G}^{(0)}(\boldsymbol{k}+\boldsymbol{Q}, m+\lambda, i\epsilon_\ell) + \sum_{\boldsymbol{k},m,i\epsilon_\ell} |\bar{\Phi}_{-\lambda}|^2\mathscr{G}^{(0)}(\boldsymbol{k}-\boldsymbol{Q}, m-\lambda, i\epsilon_\ell)\mathscr{G}^{(0)}(\boldsymbol{k}, m, i\epsilon_\ell)\right) \\
=& \sum_{\boldsymbol{k},i\epsilon_\ell,m,\lambda} 2g_{\rm T}^2|\Phi_\lambda|^2\mathscr{G}^{(0)}(\boldsymbol{k}, m, i\epsilon_\ell)\mathscr{G}^{(0)}(\boldsymbol{k}+\boldsymbol{Q}, m+\lambda, i\epsilon_\ell).
\end{aligned}$$

Thus, the quadratic contribution to the free energy becomes

$$\begin{aligned}
F^{(2)}[\bar{\Phi}, \Phi] =& \sum_{\lambda} \hbar\omega_{\boldsymbol{Q},\lambda}|\Phi_\lambda|^2 + \frac{g_{\rm T}^2}{\beta N_{\rm i}} \sum_{\boldsymbol{k},i\epsilon_\ell,m,\lambda} \mathscr{G}^{(0)}(\boldsymbol{k}, m, i\epsilon_\ell)\mathscr{G}^{(0)}(\boldsymbol{k}+\boldsymbol{Q}, m+\lambda, i\epsilon_\ell)|\Phi_\lambda|^2 \\
\equiv& \alpha_+(T)|\Phi_{+1}|^2 + \alpha_-(T)|\Phi_{-1}|^2.
\end{aligned}$$

The same result can also be obtained by evaluating the Feynman diagram shown in Fig. 4 in End Matter. Moreover, the coefficients $\alpha_\pm$ correspond to the bubble diagram within the RPA in Fig. S1, consistent with the microscopic analysis. Therefore, the quadratic term gives each transition temperature $T_{\rm C,\pm}$ of the CDW order:

$$\begin{aligned}
\alpha_\pm(T) \simeq& \hbar\omega_{\boldsymbol{Q},\pm} + \frac{g_{\rm T}^2}{\beta N_{\rm i}} \sum_{\boldsymbol{k},i\epsilon_\ell,m} \mathscr{G}^{(0)}(\boldsymbol{k}, m, i\epsilon_\ell)\mathscr{G}^{(0)}(\boldsymbol{k}+\boldsymbol{Q}, m\pm 1, i\epsilon_\ell) && \text{(S.42)} \\
=& \hbar\omega_{\boldsymbol{Q},\pm}\left[1 - \frac{Q^2 g_{\rm T}^2}{M_{\rm i}\omega_{\boldsymbol{Q},\pm}^2}\mathcal{L}_\pm(\boldsymbol{Q}, 0)\right] && \text{(S.43)} \\
\simeq& \alpha(T - T_{\rm C,\pm}). && \text{(S.44)}
\end{aligned}$$

Similarly, the quartic term can also be calculated using the Feynman diagrams shown in the End Matter; here, for completeness, we explicitly expand the expression to confirm the same result. The fourth-order term can be explicitly written as

$$\begin{aligned}
\mathrm{tr}(\mathscr{G}^{(0)}\mathcal{V})^4 = \sum_{(k_1,m_1,\lambda_1)} \cdots \sum_{(k_4,m_4,\lambda_4)} & \mathscr{G}^{(0)}(k_1,m_1)\mathcal{V}_{(k_1,m_1,\lambda_1),(k_2,m_2,\lambda_2)}\mathscr{G}^{(0)}(k_2,m_2)\mathcal{V}_{(k_2,m_2,\lambda_2),(k_3,m_3,\lambda_3)} \\
& \times\mathscr{G}^{(0)}(k_3,m_3)\mathcal{V}_{(k_3,m_3,\lambda_3),(k_4,m_4,\lambda_4)}\mathscr{G}^{(0)}(k_4,m_4)\mathcal{V}_{(k_4,m_4,\lambda_4),(k_1,m_1,\lambda_1)}.
\end{aligned} \tag{S.45}$$

The fourth-order term can be systematically classified according to combinations of the order parameters appearing in $\mathrm{tr}(\mathscr{G}^{(0)}\mathcal{V})^4$. Specifically, the contributions which satisfy the selection rule of CAM , $\lambda_1 + \lambda_2 + \lambda_3 + \lambda_4 = 0$, are grouped into the following three types:

(A) terms with a single $\lambda$: $(\bar{\Phi}_\lambda, \Phi_\lambda, \bar{\Phi}_\lambda, \Phi_\lambda)$, which yield ${}_4C_2 = 6$ distinct permutations

(B) terms with two $\lambda$: $(\bar{\Phi}_{\lambda_1}, \Phi_{\lambda_1}, \bar{\Phi}_{\lambda_2}, \Phi_{\lambda_2})$ with $\lambda_1 \neq \lambda_2$, giving $4! = 24$ permutations

(C) terms with multiple $\lambda$: $(\bar{\Phi}_{\lambda_1}, \Phi_{\lambda_2}, \bar{\Phi}_{-\lambda_3}, \Phi_{-\lambda_4})$ under the constraint $\lambda_1 + \lambda_2 = \lambda_3 + \lambda_4$, which do not arise for $\lambda = \pm 1$ in $\hat{C}_3$-symmetric systems.

Based on this classification, the contribution (A) is given by

$$\begin{aligned}(\mathrm{A}) : \sum_{\lambda} \sum_{\boldsymbol{k}, m, i\epsilon_\ell} g_T^4 \Big[ & 2\mathscr{G}^{(0)}(\boldsymbol{k}, m, i\epsilon_\ell)^2 \mathscr{G}^{(0)}(\boldsymbol{k}+\boldsymbol{Q}, m+\lambda, i\epsilon_\ell)^2 \\ & +4\mathscr{G}^{(0)}(\boldsymbol{k}, m, i\epsilon_\ell)\mathscr{G}^{(0)}(\boldsymbol{k}+\boldsymbol{Q}, m+\lambda, i\epsilon_\ell)^2 \mathscr{G}^{(0)}(\boldsymbol{k}+2\boldsymbol{Q}, m+2\lambda, i\epsilon_\ell)\Big] |\Phi_\lambda|^4,\end{aligned}$$

while the contribution (B) is given by

$$\begin{aligned}(\mathrm{B}) : \sum_{\lambda_1 \neq \lambda_2} \sum_{\boldsymbol{k}, m, i\epsilon_\ell} g_T^4 \Big[ & 8\mathscr{G}^{(0)}(\boldsymbol{k}, m, i\epsilon_\ell)^2 \mathscr{G}^{(0)}(\boldsymbol{k}+\boldsymbol{Q}, m+\lambda_1, i\epsilon_\ell)\mathscr{G}^{(0)}(\boldsymbol{k}+\boldsymbol{Q}, m+\lambda_2, i\epsilon_\ell) \\ & +8\mathscr{G}^{(0)}(\boldsymbol{k}, m, i\epsilon_\ell)\mathscr{G}^{(0)}(\boldsymbol{k}+\boldsymbol{Q}, m+\lambda_1, i\epsilon_\ell)\mathscr{G}^{(0)}(\boldsymbol{k}+2\boldsymbol{Q}, m+\lambda_1+\lambda_2, i\epsilon_\ell)\mathscr{G}^{(0)}(\boldsymbol{k}+\boldsymbol{Q}, m+\lambda_2, i\epsilon_\ell) \\ & +8\mathscr{G}^{(0)}(\boldsymbol{k}, m, i\epsilon_\ell)\mathscr{G}^{(0)}(\boldsymbol{k}+\boldsymbol{Q}, m+\lambda_1, i\epsilon_\ell)^2 \mathscr{G}^{(0)}(\boldsymbol{k}, m+\lambda_1-\lambda_2, i\epsilon_\ell)\Big] |\Phi_{\lambda_1}|^2 |\Phi_{\lambda_2}|^2.\end{aligned}$$

From these contributions, when the nesting between the $m = \pm 1$ electronic bands is taken into account, the quartic free energy takes the form

$$\mathcal{F}^{(4)}[\bar{\Phi}, \Phi] = \frac{g_{\mathrm{T}}^4}{4} I_1 (|\Phi_{+1}|^4 + |\Phi_{-1}|^4) + \frac{g_{\mathrm{T}}^4}{4} I_2 |\Phi_{+1}|^2 |\Phi_{-1}|^2,$$

where

$$\begin{aligned}I_1 \equiv \sum_{\boldsymbol{k}, i\epsilon_\ell} \frac{1}{N_{\mathrm{i}}\beta} \Big[ & 2\mathscr{G}^{(0)}(\boldsymbol{k}, +1, i\epsilon_\ell)^2 \mathscr{G}^{(0)}(\boldsymbol{k}+\boldsymbol{Q}, -1, i\epsilon_\ell)^2 \\ & +4\mathscr{G}^{(0)}(\boldsymbol{k}, +1, i\epsilon_\ell)\mathscr{G}^{(0)}(\boldsymbol{k}+\boldsymbol{Q}, -1, i\epsilon_\ell)^2 \mathscr{G}^{(0)}(\boldsymbol{k}+2\boldsymbol{Q}, 0, i\epsilon_\ell)\Big], \\ I_2 \equiv \sum_{\boldsymbol{k}, i\epsilon_\ell} \frac{1}{N_{\mathrm{i}}\beta} \Big[ & 8\mathscr{G}^{(0)}(\boldsymbol{k}, +1, i\epsilon_\ell)^2 \mathscr{G}^{(0)}(\boldsymbol{k}+\boldsymbol{Q}, -1, i\epsilon_\ell)\mathscr{G}^{(0)}(\boldsymbol{k}+\boldsymbol{Q}, 0, i\epsilon_\ell) \\ & +8\mathscr{G}^{(0)}(\boldsymbol{k}, +1, i\epsilon_\ell)\mathscr{G}^{(0)}(\boldsymbol{k}+\boldsymbol{Q}, -1, i\epsilon_\ell)\mathscr{G}^{(0)}(\boldsymbol{k}+2\boldsymbol{Q}, +1, i\epsilon_\ell)\mathscr{G}^{(0)}(\boldsymbol{k}+\boldsymbol{Q}, 0, i\epsilon_\ell) \\ & +8\mathscr{G}^{(0)}(\boldsymbol{k}, +1, i\epsilon_\ell)\mathscr{G}^{(0)}(\boldsymbol{k}+\boldsymbol{Q}, -1, i\epsilon_\ell)^2 \mathscr{G}^{(0)}(\boldsymbol{k}, 0, i\epsilon_\ell)\Big].\end{aligned}$$

Finally, the Landau free energy expanded up to the fourth order is obtained as

$$\begin{aligned}\mathcal{F} &= \mathcal{F}_0 + \alpha_+(T)|\Phi_{+1}|^2 + \alpha_-(T)|\Phi_{-1}|^2 + \beta_1(|\Phi_{+1}|^4 + |\Phi_{-1}|^4) + \beta_2 |\Phi_{+1}|^2 |\Phi_{-1}|^2, \\ \alpha_\pm &= \alpha(T - T_{C,\pm}), \\ \beta_1 &= \frac{g_T^4}{4} I_1, \\ \beta_2 &= \frac{g_T^4}{4} I_2.\end{aligned}$$

**Stable CDW order**

We now analyze the equilibrium states obtained by minimizing the Landau free energy. In the achiral crystals, each transition temperature of the LH and RH axial phonon condensation coincides $T_{\mathrm{C},+} = T_{\mathrm{C},-} \equiv T_{\mathrm{C}}$, and thus $\alpha_\pm = \alpha(T - T_{\mathrm{C}})$ near the transition temperature. The variational equations are

$$\frac{\partial F}{\partial \bar{\Phi}_{+1}} = \alpha(T - T_{\mathrm{C},+})\Phi_{+1} + 2\beta_1 |\Phi_{+1}|^2 \Phi_{+1} + \beta_2 |\Phi_{-1}|^2 \Phi_{+1} = 0, \tag{S.46}$$

$$\frac{\partial F}{\partial \bar{\Phi}_{-1}} = \alpha(T - T_{\mathrm{C},-})\Phi_{-1} + 2\beta_1 |\Phi_{-1}|^2 \Phi_{-1} + \beta_2 |\Phi_{+1}|^2 \Phi_{-1} = 0. \tag{S.47}$$

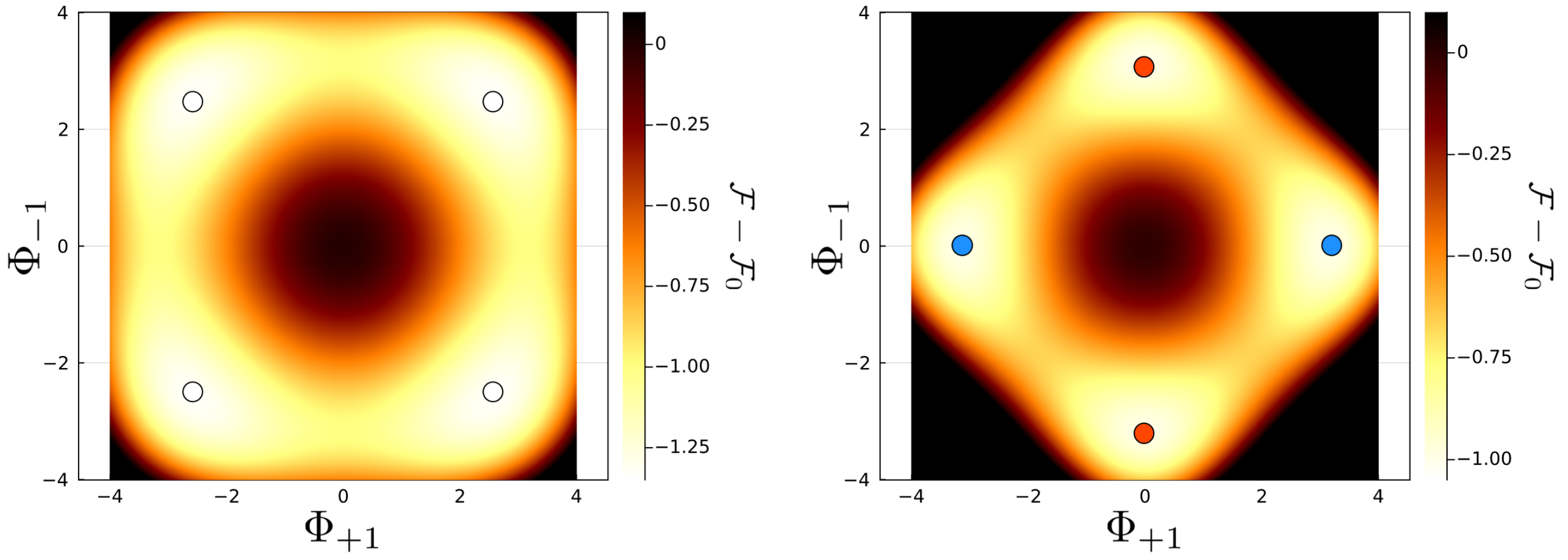


Figure S2. The heatmap of Landau free energy $\mathcal{F}-\mathcal{F}_0$ for the case (i) (left panel) and case (ii) (right panel). The white circles in the left panel mark the energy minimum $\mathcal{F}^Q_{\text{min}}$ corresponding to the nematic CDW. The red and blue circles in the right panel mark the minima $\mathcal{F}^W_{\text{min}}$ corresponding to the LH and RH winding CDWs, respectively.

For $T > T_{\text{C}}$, the only solution is the trivial one $|\Phi_{+1}| = |\Phi_{-1}| = 0$, corresponding to the normal state. Below $T_{\text{C}}$, two distinct ordered phases compete.

Let us consider the case where only a single axial phonon mode condenses, e.g., $|\Phi_{+1}| \neq 0, |\Phi_{-1}| = 0$. Substituting into the first variational equation yields the nontrivial solution

$$|\Phi_{+1}|^2 = \frac{\alpha(T_{\text{C}}-T)}{2\beta_1}. \tag{S.48}$$

The corresponding free energy minimum is

$$\mathcal{F}^W_{\text{min}} = \mathcal{F}_0 - \frac{\alpha^2(T_{\text{C}}-T)^2}{4\beta_1}. \tag{S.49}$$

By symmetry, an equivalent minimum exists at $|\Phi_{-1}| \neq 0, |\Phi_{+1}| = 0$ with the same free energy. As shown in Fig. 3(b) in the main text, the minimum at $|\Phi_{+1}| \neq 0$ corresponds to the RH winding CDW while the minimum at $|\Phi_{-1}| \neq 0$ corresponds to the LH winding CDW. The two degenerate minima are related by spontaneous breaking of the mirror symmetry of the parent achiral crystal.

On the other hand, the achiral nematic CDW is stabilized when both axial phonon modes condense with equal amplitude, $|\Phi_{+1}| = |\Phi_{-1}| \equiv |\Phi| \neq 0$. Adding the two variational equations gives

$$|\Phi|^2 = \frac{\alpha(T_{\text{C}}-T)}{2\beta_1+\beta_2}. \tag{S.50}$$

Therefore, the corresponding free energy minimum is

$$\mathcal{F}^Q_{\text{min}} = \mathcal{F}_0 - \frac{\alpha^2(T_{\text{C}}-T)^2}{2\beta_1+\beta_2}. \tag{S.51}$$

The relative stability of the two ordered phases is determined by comparing their free energy minima:

$$\mathcal{F}^Q_{\text{min}} - \mathcal{F}^W_{\text{min}} = \alpha^2(T_{\text{C}}-T)^2 \cdot \frac{\beta_2-2\beta_1}{4\beta_1(2\beta_1+\beta_2)}. \tag{S.52}$$

Since $\alpha^2(T_{\text{C}}-T)^2 > 0$ and $\beta_1, \beta_2 > 0$ in the second-order phase transition, the realized order can be classified as follows:

(i) $\beta_2 < 2\beta_1$: $\mathcal{F}^Q_{\text{min}} < \mathcal{F}^W_{\text{min}}$, and the nematic CDW is the thermodynamically stable phase.

(ii) $\beta_2 > 2\beta_1$: $\mathcal{F}^Q_{\text{min}} > \mathcal{F}^W_{\text{min}}$, and the winding CDW is the thermodynamically stable phase.

The free energy landscape is illustrated for both cases in Fig. S2. The left panel shows the case $\alpha = -0.2$, $\beta_1 = 0.01$, $\beta_2 = 0.01$, where $\beta_2 < 2\beta_1$ is satisfied and $\mathcal{F}^Q_{\rm min}$ constitutes the global minimum; here, the two order parameters condense simultaneously, realizing the nematic CDW. The right panel shows the case $\alpha = -0.2$, $\beta_1 = 0.01$, $\beta_2 = 0.04$, where $\beta_2 > 2\beta_1$ and $\mathcal{F}^W_{\rm min}$ is the global minimum. In this regime, the free energy landscape exhibits two distinct minima: stabilization at the minimum with $|\Phi_{+1}| \neq 0$ realizes the RH winding CDW while $|\Phi_{-1}| \neq 0$ realizes the LH winding CDW. The selection between these two degenerate minima is determined spontaneously, corresponding to an achiral-to-chiral phase transition.

**Evaluation of $\beta_2/\beta_1$**

To evaluate the ratio $\beta_2/\beta_1 = I_2/I_1$, we start from the band structure

$$\begin{aligned} \xi_{\boldsymbol{k},m} &= \varepsilon_m(\boldsymbol{k}) - \mu, \\ \varepsilon_0(\boldsymbol{k}) &= U_0 - t_{\parallel,0}\,\gamma(k_x,k_y) - t_0\cos(k_z c), \\ \varepsilon_{\pm 1}(\boldsymbol{k}) &= U_1 - t_{\parallel,1}\,\gamma(k_x,k_y) - t_1\cos(k_z c), \end{aligned} \tag{S.53}$$

where $\gamma(k_x, k_y)$ is the in-plane dispersion. Since we focus on the electron and phonon bands along the $z$-axis and the nesting vector $\boldsymbol{Q} = (0, 0, Q)$, we can assume that the dominant contribution to $I_1$ and $I_2$ comes from the vicinity of this line and the in-plane hopping $t_\parallel$ is sufficiently small. Therefore, the band dispersion becomes independent of $(k_x, k_y)$, so that integration over

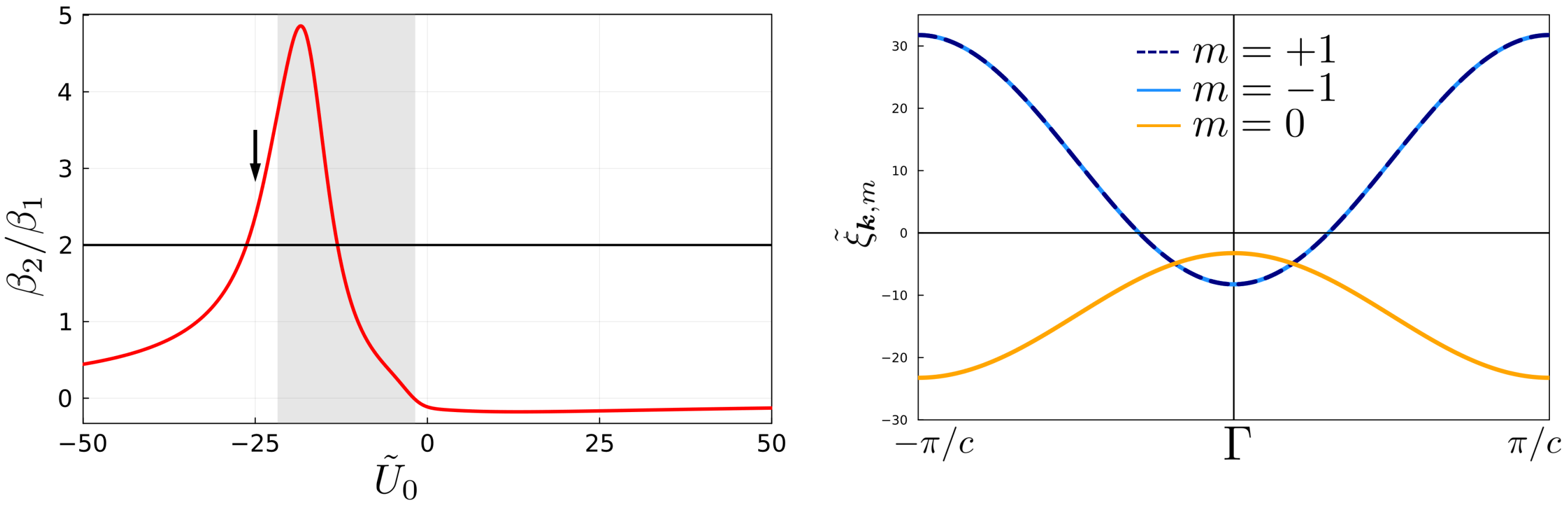

Figure S3. (Left) The ratio $\beta_2/\beta_1$ as a function of $\tilde{U}_0$ for $\tilde{t}_0 = 10$, $\tilde{t}_1 = -20$, and $k_{\rm F} = 0.3\pi/c$. The horizontal line indicates $\beta_2/\beta_1 = 2$, above which the winding CDW is stabilized. The gray rectangle represents the region where the Fermi surface of the $m = 0$ band exists. The arrow indicates the value $\tilde{U}_0 = -25$ at which the electronic bands are shown in the right panel. (Right) The electronic band structure $\tilde{\xi}_{\boldsymbol{k},m}$ at $\tilde{U}_0 = -25$. The left and right panels correspond to Fig. 3(d) and (e) in the main text, respectively.

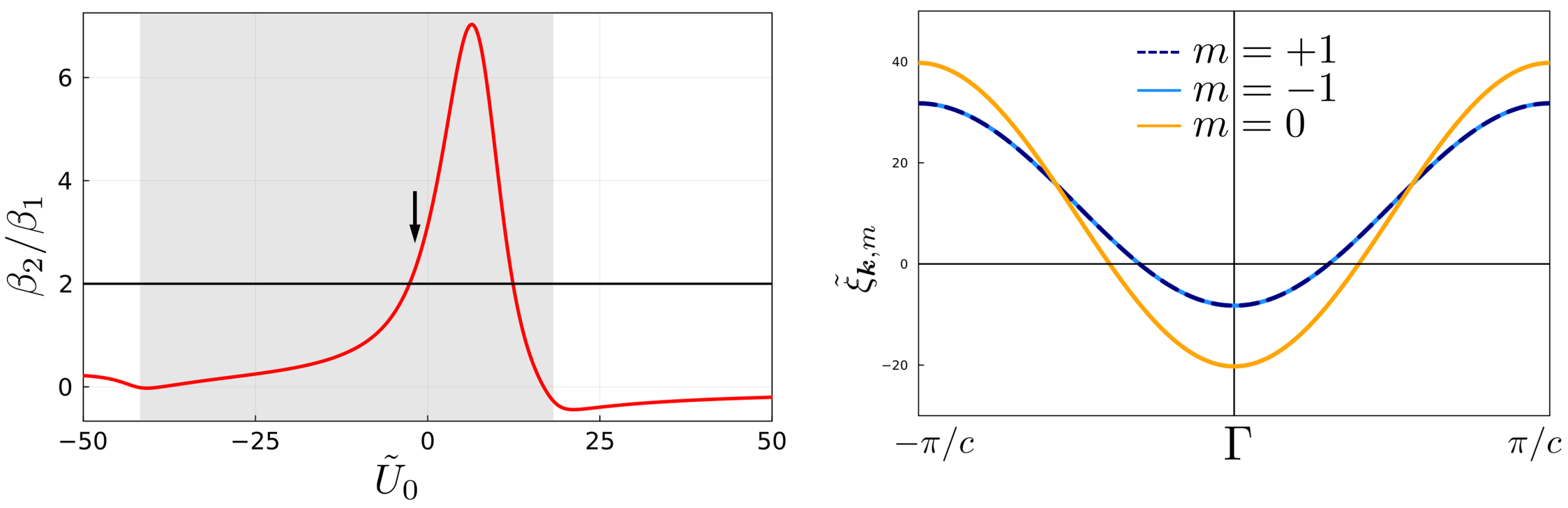

Figure S4. Same as Fig. S3 but for $\tilde{t}_0 = -30$. The electronic bands are shown at $\tilde{U}_0 = -2$.

$(k_x, k_y)$ contributes as an overall prefactor, the effective nesting area $S_\parallel$. In addition, we define the dimensionless parameters scaled by temperature as

$$\tilde{\xi}_{k_z,m} = \tilde{\varepsilon}_m(k_z) - \tilde{\mu},$$
$$\tilde{\varepsilon}_0(k_z) = \tilde{U}_0 - \tilde{t}_0 \cos(k_z c),$$
$$\tilde{\varepsilon}_{\pm 1}(k_z) = \tilde{U}_1 - \tilde{t}_1 \cos(k_z c),$$

where $\tilde{U} = U/(k_\mathrm{B}T)$, $\tilde{t} = t/(k_\mathrm{B}T)$, and $\tilde{\mu} = \mu/(k_\mathrm{B}T)$. Since we are concerned with the transverse Peierls transition driven by $l = \pm 1$ axial phonon condensation, we focus on the two nesting channels involving the coupling between electronic states near $(-k_\mathrm{F},\, m = +1)$ and $(k_\mathrm{F},\, m = -1)$, and between those near $(-k_\mathrm{F},\, m = -1)$ and $(k_\mathrm{F},\, m = +1)$, which are degenerate in achiral crystals.

The integrals $I_1$ and $I_2$ formulated in the previous section then reduce to

$$\begin{aligned} I_1 &= \int \frac{dk_x dk_y}{(2\pi)^2} \int \frac{dk_z}{2\pi} \frac{1}{\beta} \sum_{i\omega_n} \Bigg[ 2 \frac{1}{(i\omega_n - \xi_{\boldsymbol{k},+1})^2} \frac{1}{(i\omega_n - \xi_{\boldsymbol{k}+\boldsymbol{Q},-1})^2} \\ &\qquad + 4 \frac{1}{i\omega_n - \xi_{\boldsymbol{k},+1}} \frac{1}{(i\omega_n - \xi_{\boldsymbol{k}+\boldsymbol{Q},-1})^2} \frac{1}{i\omega_n - \xi_{\boldsymbol{k}+2\boldsymbol{Q},0}} \Bigg] \\ &= \frac{S_\parallel}{(k_\mathrm{B}T)^3} \int_{-\pi}^{\pi} \frac{dk_z}{2\pi} \sum_{n=-\infty}^{\infty} \Bigg[ 2 \frac{1}{[i(2n+1)\pi - \tilde{\xi}_{k_z,+1}]^2} \frac{1}{[i(2n+1)\pi - \tilde{\xi}_{k_z+Q,-1}]^2} \\ &\qquad + 4 \frac{1}{i(2n+1)\pi - \tilde{\xi}_{k_z,+1}} \frac{1}{[i(2n+1)\pi - \tilde{\xi}_{k_z+Q,-1}]^2} \frac{1}{i(2n+1)\pi - \tilde{\xi}_{k_z+2Q,0}} \Bigg], \end{aligned} \tag{S.54}$$

$$\begin{aligned} I_2 &= \int \frac{dk_x dk_y}{(2\pi)^2} \int \frac{dk_z}{2\pi} \frac{1}{\beta} \sum_{i\omega_n} \Bigg[ 8 \frac{1}{(i\omega_n - \xi_{\boldsymbol{k},+1})^2} \frac{1}{i\omega_n - \xi_{\boldsymbol{k}+\boldsymbol{Q},-1}} \frac{1}{i\omega_n - \xi_{\boldsymbol{k}+\boldsymbol{Q},0}} \\ &\qquad + 8 \frac{1}{i\omega_n - \xi_{\boldsymbol{k},+1}} \frac{1}{i\omega_n - \xi_{\boldsymbol{k}+\boldsymbol{Q},-1}} \frac{1}{i\omega_n - \xi_{\boldsymbol{k}+2\boldsymbol{Q},+1}} \frac{1}{i\omega_n - \xi_{\boldsymbol{k}+\boldsymbol{Q},0}} \\ &\qquad + 8 \frac{1}{i\omega_n - \xi_{\boldsymbol{k},+1}} \frac{1}{(i\omega_n - \xi_{\boldsymbol{k}+\boldsymbol{Q},-1})^2} \frac{1}{i\omega_n - \xi_{\boldsymbol{k},0}} \Bigg] \\ &= \frac{S_\parallel}{(k_\mathrm{B}T)^3} \int_{-\pi}^{\pi} \frac{dk_z}{2\pi} \sum_{n=-\infty}^{\infty} \Bigg[ 8 \frac{1}{[i(2n+1)\pi - \tilde{\xi}_{k_z,+1}]^2} \frac{1}{i(2n+1)\pi - \tilde{\xi}_{k_z+Q,-1}} \frac{1}{i(2n+1)\pi - \tilde{\xi}_{k_z+Q,0}} \\ &\qquad + 8 \frac{1}{i(2n+1)\pi - \tilde{\xi}_{k_z,+1}} \frac{1}{i(2n+1)\pi - \tilde{\xi}_{k_z+Q,-1}} \frac{1}{i(2n+1)\pi - \tilde{\xi}_{k_z+2Q,+1}} \frac{1}{i(2n+1)\pi - \tilde{\xi}_{k_z+Q,0}} \\ &\qquad + 8 \frac{1}{i(2n+1)\pi - \tilde{\xi}_{k_z,+1}} \frac{1}{[i(2n+1)\pi - \tilde{\xi}_{k_z+Q,-1}]^2} \frac{1}{i(2n+1)\pi - \tilde{\xi}_{k_z,0}} \Bigg]. \end{aligned}$$

Since we are interested in the ratio $\beta_2/\beta_1 = I_2/I_1$ at a fixed temperature above the transition temperature $T_\mathrm{C}$, the common factor $S_\parallel/(k_\mathrm{B}T)^3$ cancels, and the problem reduces to a one-dimensional integral along $k_z$ and the Matsubara summation.

Typical CDW transition temperatures are of the order 10–100 K, giving $k_\mathrm{B}T$ of the order 1–10 meV. Since typical hopping parameters and crystal-field splittings are of the order 0.1–1 eV, the dimensionless parameters $\tilde{t}$ and $\tilde{U}$ take values in the range 10–100 in realistic materials. We set the energy reference such that $\tilde{U}_1 - \tilde{\mu} = 0$, and place the Fermi level on the $m = \pm 1$ bands, with the $2k_\mathrm{F}$ instability occurring between the states $(-k_\mathrm{F}, m = \pm 1)$ and $(k_\mathrm{F}, m = \mp 1)$ with $Q = 2k_\mathrm{F}$.

For a first representative example, we set $\tilde{t}_0 = 10$, $\tilde{t}_1 = -20$, $k_\mathrm{F} = 0.3\pi/c$, and vary $\tilde{U}_0$ to control the position of the $m = 0$ band relative to the Fermi level. The resulting ratio $\beta_2/\beta_1$ as a function of $\tilde{U}_0$ is shown in Fig. S3, together with the electronic bands at $\tilde{U}_0 = -25$. For a second example, we set $\tilde{t}_0 = -30$, $\tilde{t}_1 = -20$, $k_\mathrm{F} = 0.3\pi/c$, and perform the same analysis; the ratio $\beta_2/\beta_1$ and the corresponding bands at $\tilde{U}_0 = -2$ are shown in Fig. S4. In both cases, $\beta_2/\beta_1$ exceeds 2 in a finite parameter range, confirming that the winding CDW is stabilized. The enhancement of $\beta_2$ relative to $\beta_1$ occurs when the $m = 0$ band lies close to the Fermi level, where it enhances the inter-channel processes encoded in $I_2$ through multiband effects, as discussed in the main text.

**Rotational phase matching for finite-$k_\perp$ nesting states**

In the main text, the PAM selection rule is derived for electronic states on the rotation axis, where the discrete angular momentum is well defined. For the CDW instability in realistic materials such as $EuAl_4$, however, the relevant electronic states are often located at finite in-plane momenta, somewhat away from the high-symmetry lines. In this situation, the PAM itself is no longer a good quantum number. Nevertheless, when the nesting geometry forms a rotationally symmetric scattering network, the electron-phonon coupling remains constrained by the relative rotational phase structure of the Bloch wave functions.

We consider the electron-phonon matrix element

$$g_{12}(\boldsymbol{k},\boldsymbol{q}) = \langle\psi_{2,\boldsymbol{k}+\boldsymbol{q}}|\delta V_{l,\boldsymbol{q}}|\psi_{1,\boldsymbol{k}}\rangle, \tag{S.55}$$

where $|\psi_{1,\boldsymbol{k}}\rangle$ and $|\psi_{2,\boldsymbol{k}+\boldsymbol{q}}\rangle$ are Bloch states belonging to bands $m_1$ and $m_2$ — or $j_1$ and $j_2$ in spinful cases— respectively, and $\delta V_{l,\boldsymbol{q}}$ is the lattice perturbation associated with a phonon carrying CAM $l$.

For the CDW instability considered here, the nesting vector is parallel to the rotation axis, $\boldsymbol{Q} = (0, 0, Q)$, so that $\hat{C}_n\boldsymbol{Q} = \boldsymbol{Q}$. Under the $n$-fold rotation operator $\hat{C}_n$, the Bloch states transform as

$$\hat{C}_n|\psi_{1,\boldsymbol{k}}\rangle = e^{-i\theta_1(\boldsymbol{k})}|\psi_{1,C_n\boldsymbol{k}}\rangle, \tag{S.56}$$

and

$$\hat{C}_n|\psi_{2,\boldsymbol{k}+\boldsymbol{q}}\rangle = e^{-i\theta_2(\boldsymbol{k}+\boldsymbol{q})}|\psi_{2,C_n(\boldsymbol{k}+\boldsymbol{q})}\rangle, \tag{S.57}$$

where $\theta_1(\boldsymbol{k})$ and $\theta_2(\boldsymbol{k}+\boldsymbol{q})$ represent the rotational phases accumulated by the corresponding Bloch states under the crystal rotation. The phonon perturbation transforms as

$$\hat{C}_n^{-1}\delta V_{l,\boldsymbol{q}}\hat{C}_n = e^{-i2\pi l/n}\delta V_{l,\boldsymbol{q}}, \tag{S.58}$$

reflecting the CAM carried by the phonon mode.

Using these relations, the matrix element at the rotated momentum satisfies

$$g_{12}(C_n\boldsymbol{k}, C_n\boldsymbol{Q}) = g_{12}(C_n\boldsymbol{k},\boldsymbol{Q}) = e^{i[\theta_2(\boldsymbol{k}+\boldsymbol{Q})-\theta_1(\boldsymbol{k})-2\pi l/n]}g_{12}(\boldsymbol{k},\boldsymbol{Q}). \tag{S.59}$$

Moreover, the nested Fermi surface patches are related to each other by rotational symmetry. In $EuAl_4$, for example, the relevant nested side surfaces form a $\hat{C}_4$-symmetric set around the $k_z$ axis. Consequently, the scattering processes $\boldsymbol{k} \to \boldsymbol{k}+\boldsymbol{Q}$ and $C_n\boldsymbol{k} \to C_n\boldsymbol{k}+\boldsymbol{Q}$ belong to the same rotationally equivalent nesting channel. In the corresponding low-energy patch description, i.e., where $\boldsymbol{k}_\perp$ is sufficiently small, the electron-phonon coupling strength is approximately invariant under the rotation,

$$g_{12}(C_n\boldsymbol{k},\boldsymbol{Q}) \simeq g_{12}(\boldsymbol{k},\boldsymbol{Q}) \simeq g_{\mathrm{T}}. \tag{S.60}$$

Under this condition, Eq. (S.59) requires

$$e^{i[\theta_2(\boldsymbol{k}+\boldsymbol{Q})-\theta_1(\boldsymbol{k})-2\pi l/n]} = 1, \tag{S.61}$$

which yields the rotational phase-matching condition,

$$\theta_2(\boldsymbol{k}+\boldsymbol{Q}) - \theta_1(\boldsymbol{k}) - \frac{2\pi l}{n} = 2\pi p, \tag{S.62}$$

where $p$ is an integer. This condition reduces to the PAM selection rule on the rotation axis. Indeed, when the CAM $j_i$ is well defined, the rotational phases become

$$\theta_i = \frac{2\pi j_i}{n}, \tag{S.63}$$

and the above equation reduces to

$$j_1 + l = j_2 \mod n, \tag{S.64}$$

which is the selection rule derived in the main text.

Therefore, although the PAM itself is not strictly defined away from the rotation axis, the PAM selection rule derived on the high-symmetry line can be naturally extended to the finite $\boldsymbol{k}_\perp$ nesting channels relevant to $EuAl_4$, where the nesting vector parallel to the rotation axis generates rotationally equivalent scattering channels related by the $\hat{C}_4$ symmetry.

---

* asano.shun.57x@st.kyoto-u.ac.jp

† yanase@scphys.kyoto-u.ac.jp


\fi

\end{document}